\documentclass[%
superscriptaddress,
 amsmath,amssymb,
 aps, 
pra,
twocolumn
]{revtex4-2}
\usepackage{graphicx} 
\usepackage{dcolumn}
\usepackage{bm}
\usepackage{hyperref}
\usepackage{physics}
\usepackage{subcaption}
\usepackage{adjustbox}
\usepackage[T1]{fontenc}
\usepackage{xcolor}
\usepackage{placeins}
\usepackage[justification=raggedright,singlelinecheck=true]{caption}

\usepackage{newfloat} 
\DeclareFloatingEnvironment[
    name=Algorithm,
    placement=tbp, 
    listname={List of Algorithms}
]{algorithm}

\usepackage{float}
\floatstyle{ruled}
\restylefloat{algorithm}

\usepackage{ragged2e}

\begin{document}

\title{Quantum Machine Learning for State Tomography Using Classical Data}

\author{Shabnam Jabeen}
\email{jabeen@umd.edu}
\affiliation{University of Maryland, College Park}
\affiliation{UMD National Quantum Lab}
\author{Dmytro Kurdydyk}
\affiliation{Davidson College}
\author{Aadi Palnitkar}
\affiliation{University of Maryland, College Park}
\author{Mihir Talati}
\affiliation{University of Maryland, College Park}
\author{Jeffrey Yan}
\affiliation{University of Maryland, College Park}
\author{Jinghong Yang}
\email{yangjh@umd.edu}
\affiliation{University of Maryland, College Park}
\affiliation{UMD National Quantum Lab}

\begin{abstract}
Reconstructing quantum states from measurement data represents a formidable challenge in quantum information science, especially as system sizes grow beyond the reach of traditional tomography methods. While recent studies have explored quantum machine learning (QML) for quantum state tomography (QST), nearly all rely on idealized assumptions, such as direct access to the unknown quantum state as quantum data input, which are incompatible with current hardware constraints. In this work, we present a QML-based tomography protocol that operates entirely on classical measurement data and is fully executable on noisy intermediate-scale quantum (NISQ) devices. Our approach employs a variational quantum circuit trained to reconstruct quantum states based solely on measurement outcomes. 
We test the method in simulation, achieving high-fidelity reconstructions of diverse quantum states, including GHZ states, spin chain ground states, and states generated by random circuits.
The protocol is then validated on quantum hardware from IBM and IonQ. Additionally, we demonstrate accurate tomography is possible using incomplete measurement bases, a crucial step towards scaling up our protocol. Our results in various scenarios illustrate successful state reconstruction with fidelity reaching 90\% or higher. To our knowledge, this is the first QML-based tomography scheme that has been implemented on real quantum processors using exclusively classical measurements. This work establishes the feasibility of QML-based tomography on current quantum platforms and offers a scalable pathway for practical quantum state reconstruction.

\end{abstract}

\maketitle

\section{Introduction}

In quantum information science, it is of critical importance to extract information from quantum systems. In particular, it is often necessary to identify and characterize a quantum state produced in experiment~\cite{madzik_precision_2022,farooq_robust_2022,walther_experimental_2005,kok_linear_2007,barreiro_generation_2005,dong_quantum_2010,dariano_quantum_2002,baur_benchmarking_2012,haffner_scalable_2005,naghiloo_quantum_2019}. Yet, this task is a nontrivial endeavor. Unlike classical systems, measuring a quantum system can ``collapse'' its wave function, changing the state it is in and destroying remaining information~\cite{griffiths_introduction_2019,sakurai_modern_2020,wootters_single_1982,dariano_impossibility_1996}. Thus, to reconstruct the initial state, a carefully designed protocol, measuring the state multiple times and in different measurement settings, is required.
This gives rise to the subject of quantum state tomography (QST)~\cite{vogel_determination_1989}, which seeks the answer to the following question: given an ensemble of identically prepared quantum states and the ability to perform measurements on them, how do we reconstruct a representation of the state based on the measurement data? For this purpose, various methods have been developed, most notably the linear inversion method and the maximum likelihood estimation (MLE)~\cite{vogel_determination_1989,hradil_quantum_1997,banaszek_maximumlikelihood_1999,james_measurement_2001}.

Quantum state tomography has been experimentally performed on small systems~\cite{haffner_scalable_2005,resch_full_2005,madzik_precision_2022,stricker_experimental_2022,hu_experimental_2024,song_10qubit_2017}; however, as the system size increases, traditional QST methods face a daunting challenge. That is, the dimension of the Hilbert space grows exponentially as the system size. 
Thus, to fully characterize a quantum state, an exponential number of parameters is needed for the density matrix or, in the case of a pure state, the state vector.
As a consequence, for QST in general, the number of measurements needed and the required computation resource for data processing would scale up exponentially~\cite{banaszek_focus_2013,toninelli_concepts_2019}.

To overcome this problem, people have been exploring various options \cite{aaronson_shadow_2018,huang_predicting_2020,toth_permutationally_2010,moroder_permutationally_2012,gross_quantum_2010,gaikwad_gradientdescent_2025,cramer_efficient_2010,zhao_experimental_2017,lanyon_efficient_2017,kuzmin_learning_2024,wang_scalable_2020,li_efficient_2023,sunada_efficient_2024,gomez_reconstructing_2022,akhtar_scalable_2023,teng_learning_2024,torlai_neuralnetwork_2018,smith_efficient_2021,carrasquilla_reconstructing_2019,zhang_efficient_2022,singh_maximal_2025,kokaew_bootstrapping_2024,neugebauer_neural_2020,melkani_eigenstate_2020,li_experimental_2024,cha_attentionbased_2022,wei_neuralshadow_2024,xin_localmeasurementbased_2019,schmale_efficient_2022,ahmed_quantum_2021,gaikwad_neural_2024,liu_variational_2020, hai_universal_2023, xin_quantum_2020, innan_quantum_2024, yao_quantum_2024, xiao_reconstructing_2022, gupta_variational_2022, lee_learning_2018,lee_quantum_2021}.
Notable examples include shadow tomography~\cite{aaronson_shadow_2018,huang_predicting_2020}, which extracts partial information of quantum systems using polynomial resources; permutation-invariant QST~\cite{toth_permutationally_2010,moroder_permutationally_2012}, which utilizes permutation symmetry; and compressed sensing QST~\cite{gross_quantum_2010}, which alleviates scalability challenge for low-rank density matrices. 
Other approaches adopt ansatze that have a polynomial number of parameters;
while not meant for arbitrary states, these methods are designed to work under reasonable physical assumptions.
As highlighted in \cite{fannes_finitely_1992,perez-garcia_matrix_2007,vidal_efficient_2003,hastings_solving_2006,schollwoeck_densitymatrix_2011,cramer_efficient_2010,lanyon_efficient_2017}, tensor networks can efficiently represent certain classes of quantum states, notably low-entanglement states, and thus can be used for efficient tomography~\cite{cramer_efficient_2010,zhao_experimental_2017,lanyon_efficient_2017,kuzmin_learning_2024,wang_scalable_2020,li_efficient_2023,sunada_efficient_2024,gomez_reconstructing_2022,akhtar_scalable_2023,teng_learning_2024}. 
Likewise, machine learning methods are also explored for this purpose. 
Thanks to the ability of neural networks to represent certain quantum states~\cite{carleo_solving_2017,gao_efficient_2017,huang_neural_2021,sharir_neural_2022,deng_quantum_2017,chen_equivalence_2018}, Refs.~\cite{torlai_neuralnetwork_2018,smith_efficient_2021,carrasquilla_reconstructing_2019,zhang_efficient_2022,singh_maximal_2025,kokaew_bootstrapping_2024,neugebauer_neural_2020,melkani_eigenstate_2020,li_experimental_2024,cha_attentionbased_2022,wei_neuralshadow_2024,xin_localmeasurementbased_2019,schmale_efficient_2022,ahmed_quantum_2021,gaikwad_neural_2024} use neural networks for QST.

Since quantum state tomography is an inherently quantum problem, it is natural to ask whether the power of quantum computing can be harnessed for efficient reconstruction of a quantum state. In this spirit, quantum machine learning methods (QML) for QST have received attention in recent years, and Refs.~\cite{liu_variational_2020, hai_universal_2023, xin_quantum_2020, innan_quantum_2024, yao_quantum_2024, xiao_reconstructing_2022, gupta_variational_2022, lee_learning_2018,lee_quantum_2021} have begun exploring this topic. Quantum circuits with tunable parameters, referred to as variational quantum circuits (VQC), are usually used as the ansatz to parametrize the unknown quantum state.
These results have demonstrated the potential of QML-based approaches in QST. 
However, in these works, the input to the quantum algorithm is typically the unknown target state itself, which means it is assumed the unknown quantum state can be directly loaded into the quantum computer. For current quantum hardware, this can not yet be achieved. Instead, we would like to develop a protocol where the QML algorithm only depends on the classical measurement data of the unknown state. To achieve this purpose, we modify the training procedure to accommodate classical measurement data. In this way, the algorithm can be implemented and tested in a practical setting on noisy intermediate-scale quantum (NISQ) devices.

The exploration of QML for QST may not only benefit QST, but also offer meaningful insights into QML. QML is naturally suited for quantum input data~\cite{huang_quantum_2022}; yet, such applications can be hard to realize in the near future due to hardware limitation. For classical data, it is still an open question whether and where near-future QML algorithms can offer a quantum advantage~\cite{bowles_better_2024}. The problem of QST offers a setting where the data are stored in a classical format but are generated by a quantum process. This can act as a testbed for the power of QML.

The organization of this paper is as follows.
Section~\ref{sec:background} begins with an introduction to the concept of VQC, followed by an overview of QST and a simple review of the existing literature. In Section~\ref{sec:model}, we introduce our model and methodology. 
Our results are presented in Section~\ref{sec:results}. We first confirm the validity of our method by reconstructing a variety of 3- and 6-qubit quantum states on ideal simulators, as shown in Section~\ref{sec:results:ideal}. Then, in Section~\ref{sec:results:qpu}, to demonstrate compatibility with NISQ devices, we test our method on IBM's and IonQ's quantum processors, where a variational circuit was trained to reconstruct a 3-qubit Greenberger–Horne–Zeilinger (GHZ) state. 
The above results serve as a proof of concept, but to scale up to larger systems, a reduction in the number of measurements will be needed. We explore this possibility in Section~\ref{sec:results:incomplete}, where we reconstruct quantum states with incomplete measurement bases.
Finally, in Section \ref{sec:conclusion}, we conclude the paper.

\section{Background \label{sec:background}}
\subsection{Variational quantum circuit}
\begin{figure}[h]
    \centering
    \includegraphics[width=\linewidth]{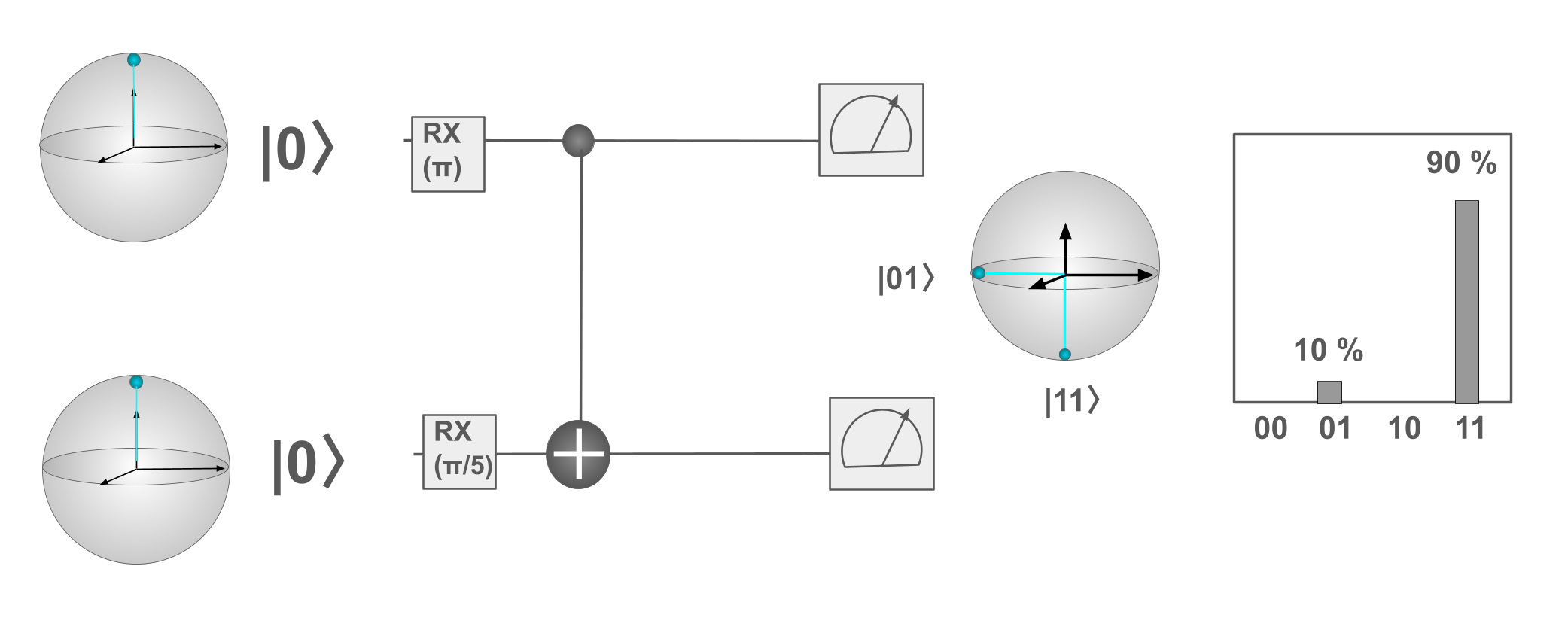}
    \caption{Example of a simple quantum circuit.}
    \label{fig:basic_quantum_circuit}
\end{figure}

The quantum circuit is the predominant quantum computing paradigm~\cite{nielsen_quantum_2000}. 
A quantum circuit consists of qubits and quantum gates (see Fig.~\ref{fig:basic_quantum_circuit}). Each qubit is a two-level quantum system, and a quantum gate is some unitary action on one or more qubits. At the end of the circuit, the circuit is measured. The measurement is often repeated a number of times (referred to as the number of shots) to obtain the expectation value of an observable.
As shown in the figure, some gates are parametrized, e.g., the $R_x$ gate being controlled by a rotation angle $\theta$.

By using parametrized gates as the trainable part of the circuit, one can construct a variational quantum circuit, thus constructing a quantum machine learning model~\cite{cerezo_variational_2021, romero_variational_2019,schuld_circuitcentric_2020,chen_variational_2020}. This is a hybrid algorithm where the quantum circuit is evaluated on a quantum device but the optimization of the parameters is handled classically. For each iteration, the circuit is executed according to the parameters. After circuit execution, the measurement data go through post-processing, and the loss function can thus be obtained. The loss function and the parameters are passed to a classical optimizer; it could be gradient-free methods, such as COBYLA~\cite{powell_direct_1998} and SPSA~\cite{spall_overview_1998}, or gradient-descent based methods. The optimizer will determine how to update the parameters for the next iteration.

\subsection{Quantum state tomography}
In many quantum experiments, it is important to determine the exact quantum state that has been prepared~\cite{dariano_quantum_2002,baur_benchmarking_2012,haffner_scalable_2005}.
However, compared to classical systems, the accurate determination of a quantum system is nontrivial.
For a quantum system, each measurement will project the quantum state probabilistically to a measurement basis state; the measurement result only reflects which basis state has been projected into, and other information is destroyed~\cite{griffiths_introduction_2019,sakurai_modern_2020,wootters_single_1982,dariano_impossibility_1996}.
Furthermore, even if the probability distribution for a certain measurement basis is exactly known, it is not enough to fully reconstruct a state, since it does not capture the coherence and relative phase information~\cite{griffiths_introduction_2019}.
To fully characterize a state, quantum state tomography (QST) is needed. 
In QST, it is assumed many copies of the unknown states can be prepared by repeating the same preparation procedure. The copies can be measured in multiple different measurement bases. 
Based on the measurement data, QST seeks to reconstruct a representation of the quantum state, most often a density matrix representation.

Many techniques have been used for QST. The most straightforward method is the \textit{linear inversion method}~\cite{vogel_determination_1989,james_measurement_2001}. The unknown state will be measured in many bases, so that the expectation value for a chosen set of observables can be measured. The set of observables should be tomographically complete, e.g., the set of n-qubit Pauli strings for an n-qubit system. Then, the measurement results can be inverted to yield the density matrix of the quantum state. One drawback of this method is that the density matrix thus obtained is not guaranteed to be positive semi-definite in the presence of noise, and it may end up having negative eigenvalues~\cite{hradil_quantum_1997}.
An alternative approach which circumvents this issue is the \textit{Maximum Likelihood Estimation} (MLE)~\cite{hradil_quantum_1997,banaszek_maximumlikelihood_1999,james_measurement_2001}. This approach reconstructs a density matrix by finding the density matrix with the maximum likelihood of producing the measurement data, typically using an iterative algorithm. Additionally, the \textit{Bayesian method} has also been investigated~\cite{blume-kohout_optimal_2010,granade_practical_2016,lukens_practical_2020}.  
All of the aforementioned approaches share one shortcoming: as the system size grows, the dimension of the Hilbert space grows exponentially. For an n-qubit system, the density matrix would be a $2^n$ by $2^n$ complex matrix. Hence, the number of measurements and the amount of computation resource needed to determine a density matrix would scale up exponentially with system size. Thus, while QST can be performed for small systems, performing it on larger systems would be prohibitively expensive.

To address this issue, many techniques are being developed, many of which resort to representations of quantum states with only a polynomial number of parameters, as opposed to the density matrix representation.
This is hinged on the concept that while a generic quantum state requires an exponential number of coefficients to describe, many quantum states of physical interest can be characterized with far fewer parameters. 
One such technique is to use tensor networks to represent certain quantum states. For example, the ground state of the 1-dimensional gapped local Hamiltonian can be efficiently represented by a type of tensor network called the matrix product state (MPS)~\cite{hastings_area_2007,verstraete_matrix_2006,cirac_matrix_2021}. The use of tensor networks in QST has been explored by \cite{cramer_efficient_2010,zhao_experimental_2017,lanyon_efficient_2017,kuzmin_learning_2024,wang_scalable_2020,li_efficient_2023,sunada_efficient_2024,gomez_reconstructing_2022,akhtar_scalable_2023,teng_learning_2024}.
The neural network is another popular choice. Motivated by studies that show the power of neural networks at representing quantum states in principle~\cite{carleo_solving_2017,gao_efficient_2017,huang_neural_2021,sharir_neural_2022,deng_quantum_2017,chen_equivalence_2018}, many studies in recent years have chosen to use a neural network to parametrize an unknown state.
The restricted Boltzmann machine (RBM) has been employed in Ref.~\cite{torlai_neuralnetwork_2018,smith_efficient_2021,carrasquilla_reconstructing_2019,zhang_efficient_2022,singh_maximal_2025,kokaew_bootstrapping_2024,neugebauer_neural_2020,melkani_eigenstate_2020}; other generative models, such as the recurrent neural network~\cite{carrasquilla_reconstructing_2019, li_experimental_2024}, transformers~\cite{cha_attentionbased_2022,wei_neuralshadow_2024,xin_localmeasurementbased_2019}, etc.~\cite{schmale_efficient_2022,ahmed_quantum_2021,gaikwad_neural_2024}, have also been used.

While the methods mentioned above utilize classical architectures for efficient quantum state representation, work in recent years has begun investigating the idea of using quantum circuits to represent quantum states for tomography purposes.
While quantum states in general require an exponential number of gates to obtain, many physically relevant states can be obtained or approximated by quantum circuits with polynomial depth~\cite{poulin_quantum_2011,ward_preparation_2009,zalka_simulating_1998,jordan_quantum_2012}.
Hence, variational quantum circuits can potentially serve as powerful ansatze for quantum states, including certain highly entangled states, with number of parameters scaling polynomially as the number of qubits.
Such techniques have been applied to QST \cite{liu_variational_2020, hai_universal_2023, xin_quantum_2020, innan_quantum_2024, yao_quantum_2024, xiao_reconstructing_2022,gupta_variational_2022,lee_learning_2018,lee_quantum_2021}, and some works have taken one step further to explore its use in quantum process tomography \cite{xue_variational_2022, xue_variational_2023, galetsky_optimal_2024, yang_active_2024}.
In these works, the following procedure or an equivalent one is typically adopted. A variational quantum circuit is used as the ansatz, and the parameters are varied to maximize the quantum fidelity between the ansatz state and the unknown quantum state, pushing the ansatz state closer to the target state.
For general states $\rho$ and $\sigma$, the quantum fidelity is defined as~\cite{jozsa_fidelity_1994}
\begin{equation*}
    F\left(\rho, \sigma\right) = \left(\text{Tr}\sqrt{\sqrt{\rho}\sigma\sqrt{\rho}}\right)^2,
\end{equation*}
and for pure states $\ket{\psi}$ and $\ket{\phi}$, the above expression can be simplified to
\begin{equation*}
    F\left(\ket{\psi},\ket{\phi}\right)=\left|\bra{\psi}\ket{\phi}\right|^2.
\end{equation*}
It is a number between $0$ and $1$.
In short, the quantum fidelity can measure the overlap between two states, and maximizing it drives the ansatz state to resemble the unknown target state.
This can be viewed as a quantum machine learning (QML) problem where the model, i.e. the variational quantum circuit, is trained to optimize a loss function.

However, while quantum fidelity is a good measure of the proximity between quantum states, its use in QST tasks in the near future can face experimental limitations. 
As mentioned in the QML-for-QST literature~\cite{liu_variational_2020, hai_universal_2023, xin_quantum_2020, innan_quantum_2024, yao_quantum_2024, xiao_reconstructing_2022, gupta_variational_2022, lee_learning_2018,lee_quantum_2021}, the quantum fidelity between the ansatz state and the unknown quantum state can be obtained by the SWAP test~\cite{buhrman_quantum_2001} or other procedures. Doing so would require both the unknown state and the ansatz state to be on the quantum computer. However, for tomography tasks~\footnote{Here, the discussion is about quantum state tomography tasks. If, instead, the goal is to approximately compress a known circuit to a shorter circuit, then quantum fidelity can be used as the loss function in practical settings.}, a more likely scenario is as follows. An unknown quantum state is produced via some procedure on a quantum device $\mathcal{A}$, where universal quantum computation cannot be reliably performed. In the meantime, there exists a reliable quantum computer $\mathcal{C}$, where one can run the variational quantum circuit and produce the ansatz state. 
In this scenario, one cannot trivially compute the fidelity between the unknown state and the ansatz state, unless quantum teleportation can be performed between $\mathcal{A}$ and $\mathcal{C}$, which cannot yet be implemented on current quantum computers. To perform such tomography in the near future, one needs to make measurements of the unknown state on device $\mathcal{A}$; then, based on the measurement results, the state can be reconstructed on quantum computer $\mathcal{C}$ via a variational algorithm. 
Our goal is to implement such a protocol, relying on classical measurement data rather than quantum data input (see Fig.~\ref{fig:qst-figure}).

Among the QML-for-QST literature, there are two exceptions which are compatible with classical data. 
In \cite{innan_quantum_2024}, one of the methods takes classical data as input; however, the quantum circuits there are not used to parametrize the quantum state directly, but are trained to generate the coefficients of the quantum state instead, similar to \cite{torlai_neuralnetwork_2018}.
On the other hand, in \cite{gupta_variational_2022}, the quantum state is parametrized as a Gibbs-like state $\rho=\exp(-\beta H)/\Tr[\exp(-\beta H)]$ as in the maximal entropy formalism. The operator $H$ is varied to match the measurement data, while a variational quantum circuit is used to produce the Gibbs-like state $\rho$ for a given $H$.
This formalism works with classical data, but has not been tested in the practical setting with finite number of shots.
Additionally, it is worthwhile to explore methods other than the maximal entropy formalism, as different approaches offer different strengths.
Therefore, a new work on using variational quantum circuits for QST with classical data is needed.

In this study, we are presenting a QML-based protocol that uses a variational quantum circuit trained solely on classical measurement data.
Rather than using the fidelity, we use loss functions amenable to classical measurement data, eliminating the need for direct quantum data input and enabling us to implement the protocol on NISQ devices.
Our methodology is detailed in the next section.

\section{Methodology \label{sec:model}}
We wish to use a quantum circuit to reconstruct an unknown quantum state. In this work, we will focus on the tomography for the pure state. 
We assume the unknown state can be measured in different Pauli bases. A possible corresponding physical scenario is where single-qubit manipulations, but not necessarily multi-qubit gates, can be reliably performed on the quantum device with the unknown state. 

\begin{figure}[htbp]
    \centering
   
    \begin{adjustbox}{minipage={0.9\linewidth},valign=b}
        \includegraphics[width=0.9\linewidth]{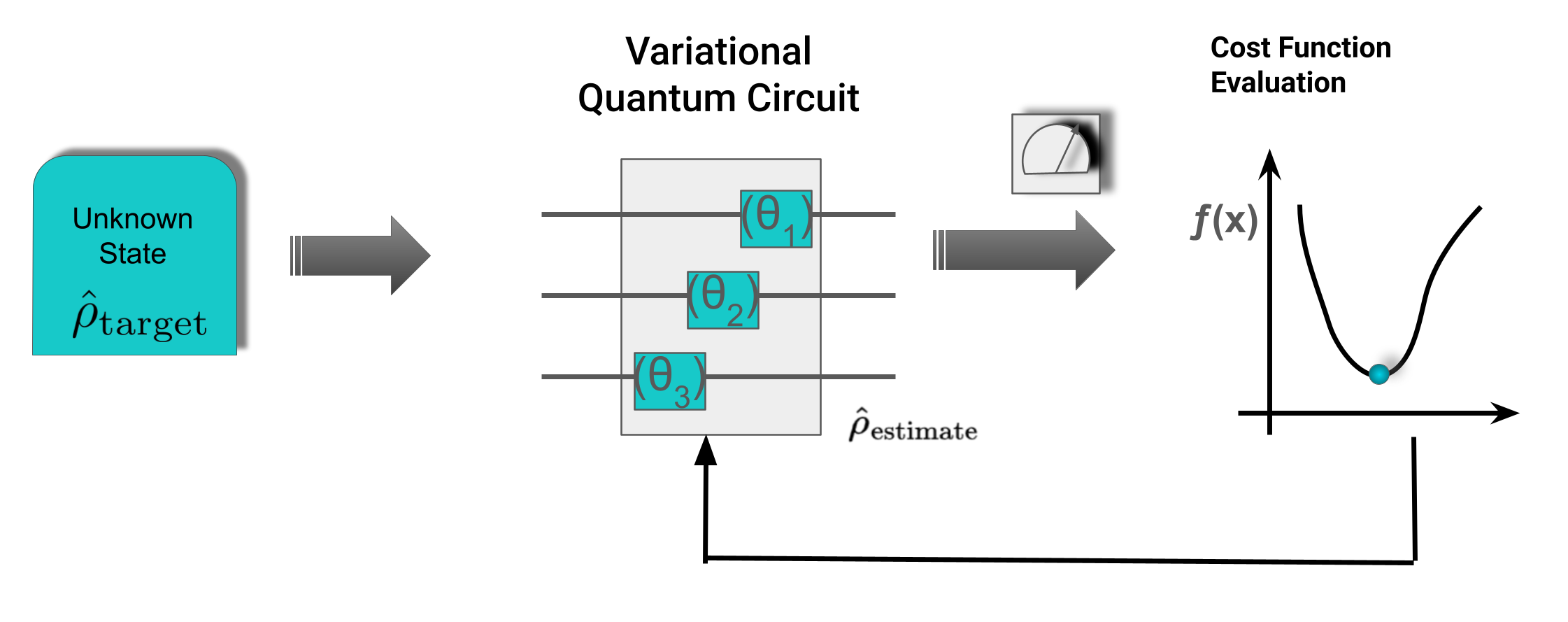}
    \end{adjustbox}
    \begin{minipage}[]{0.08\linewidth}
        \subcaption{}\label{fig:qst-figure:old}
    \end{minipage}

    \begin{adjustbox}{minipage={0.9\linewidth},valign=b}
        \includegraphics[width=\linewidth]{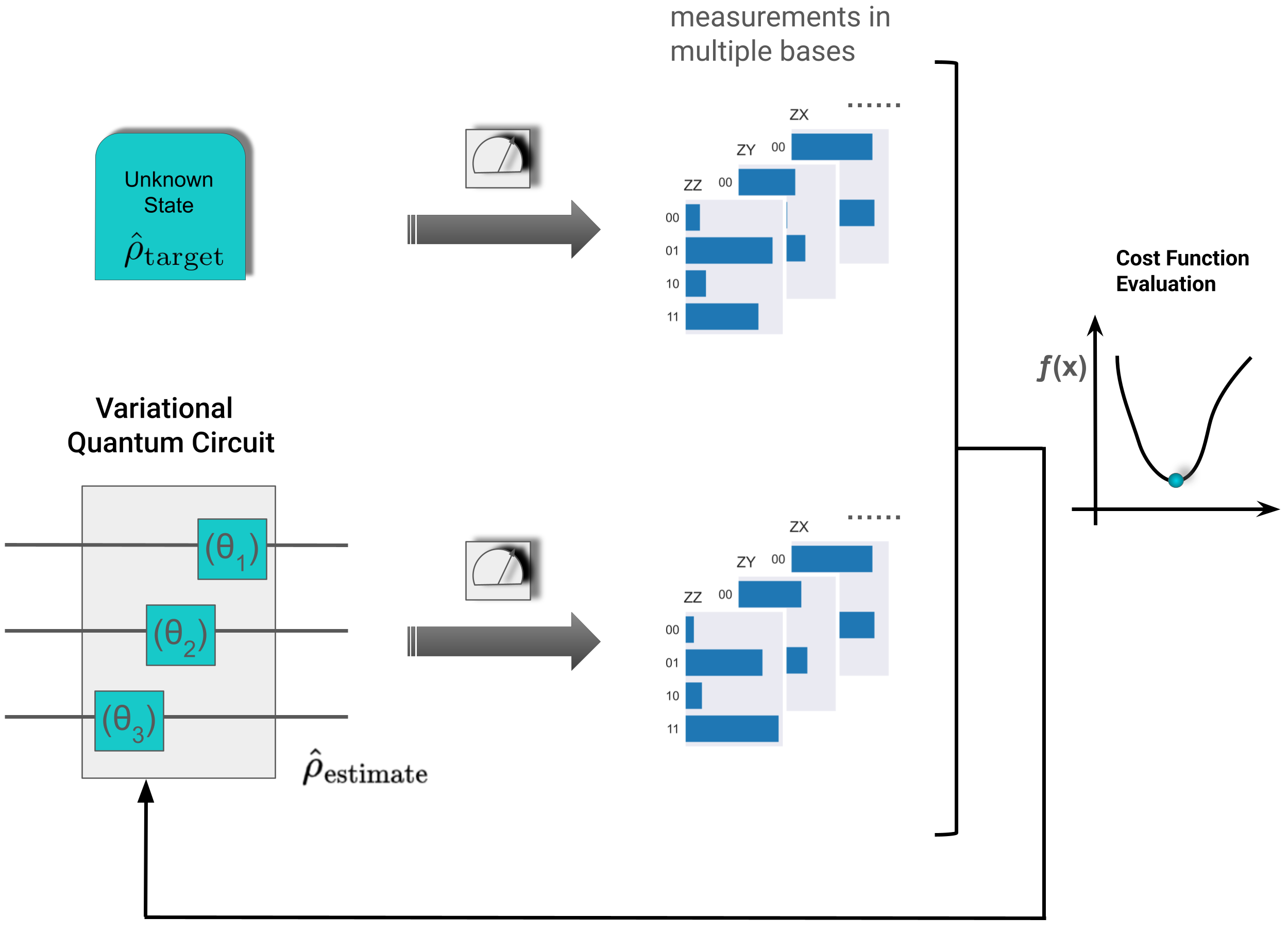}
    \end{adjustbox}
    \begin{minipage}[]{0.08\linewidth}
        \subcaption{}\label{fig:qst-figure:this}
    \end{minipage}
   
    \caption{Illustration of two QST training procedures: (a) direct quantum state input to the QML circuit; (b) the approach in this study, where the model is trained to minimize the discrepancy between the measurement data of the unknown state and that of the VQC ansatz.}
    \label{fig:qst-figure}
\end{figure}

The workflow is as described below (also see Fig.~\ref{fig:qst-figure:this} and Algorithm~\ref{alg:qst}). There is an unknown state (hereafter referred to as the ``target state'') which one wishes to reconstruct. 
A number of bases are chosen; for each qubit, the measurement can be performed in the X, Y, or Z bases. The target state is measured in these bases. 
To reconstruct the target state on a quantum circuit, we use an ansatz for a variational circuit; the state corresponding to the ansatz will hereafter be referred to as the ``ansatz state.'' The variational circuit is measured in the same bases as the target state.
Then, a certain loss function quantifies the difference between the ansatz state measurement data and that of the target state. An optimization algorithm then attempts to minimize the loss function.
At the end of the training, the ansatz state is supposed to approximately reconstruct the target state.
In the language of machine learning, this can be viewed as a generative quantum machine learning task. The measurement results of the target state are the training dataset, while the variational quantum circuit is the machine learning model, and is trained to ``generate'' data that resemble the samples in the training set.

\begin{algorithm}[htbp]
\caption {Quantum state reconstruction with variational circuit} \label{alg:qst}

\begin{minipage}{\linewidth}
\RaggedRight
\textbf{Input}: Measurement data from the unknown target state in a collection of bases $\{\vb{\sigma}^b\}$.

\textbf{Initialize}: Parameters of the variational quantum circuit are initialized as $\vec \theta_0$.

\textbf{For each iteration}

{
\leftskip=2em

\textbf{Step 1: measurement}

Parameters of the variational quantum circuit are set to be $\vec \theta$.

For each basis, measure the VQC.

\textbf{Step 2: compute loss function}

For each basis, compare the measurement data of the target state and VQC.

Quantify their discrepancy using a chosen mathematical function.

Average (or sum) over the contribution from the measured bases to obtain the loss function.

\textbf{Step 3: update parameters}

Use a classical optimizer to determine the update strategy of the parameters: $\vec\theta_\text{old}\rightarrow \vec \theta_\text{new}$.

}
\textbf{Output}: The optimal parameters $\vec \theta$ for the ansatz to reconstruct the target state.
\end{minipage}
\end{algorithm}

\begin{figure}[htbp]
    \centering
    \includegraphics[width=1\linewidth]{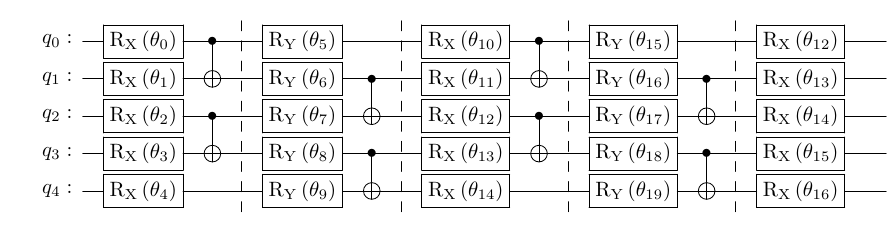}
    \caption{The circuit ansatz. The trainable layers are the alternating $R_x$ and $R_y$ layers. Between the $R_x$ and $R_y$ layers, CNOT gates are used to create entanglement. The plot shows an ansatz with 5 layers. }
    \label{fig:ansatz}
\end{figure}

For the circuit ansatz, we adopt the design from Ref.~\cite{liu_variational_2020} (see Fig.~\ref{fig:ansatz}).

For the loss function, one needs to choose a function that can measure the distance between two classical distributions to evaluate the discrepancy between the measurement data of the target state and of the ansatz state.
Let's use $b$ to denote a particular measurement basis; a measurement outcome can be denoted by a binary string $\boldsymbol{\sigma}^b$ (for example, in 2 qubit case, it can be 00, 01, 10, or 11)\footnote{For simplicity, we do not use the positive operator value measurement (POVM) formalism, but work directly with the ``raw'' measurement outcomes. The approach we use here is equivalent to Pauli-6 POVM.}. Basically, for each measurement basis, the result is characterized by the probability distribution of the binary strings $P(\boldsymbol{\sigma}^b)$. Let's denote the probability distribution of the target state by $Q^b$, and denote that of the ansatz state by $P_{\vec\theta}^b$.  To choose a loss function, we need a function $f\left(Q^b, P_{\vec\theta}^b\right)$ such that $f\left(Q^b, P_{\vec\theta}^b\right)$ is minimized when $P_{\vec\theta}^b$ is the same as $Q^b$, and its value should be large when  $P_{\vec\theta}^b$ deviates away from $Q^b$.
Here, we use the KL-divergence~\cite{shlens_notes_2014} to quantify the discrepancy between two sets of measurement data. 
We calculate the KL divergence between the probability distributions of the target state measurements and that of the ansatz state measurements for each basis and then average over different bases.
In other words, the total divergence between the sets of measurement data $P_1$ and $P_2$ is specified as
\begin{equation}
    \Xi(P_1, P_2) = \frac{1}{N_b}\sum_{b} \sum_{\boldsymbol{\sigma}^b} P_1^b(\boldsymbol{\sigma}^b) \log \frac{P_1^b(\boldsymbol{\sigma}^b)}{P_2^b(\boldsymbol{\sigma}^b) + \epsilon},
    \label{eqn:kl}
\end{equation}
where we add a small positive $\epsilon$ in the denominator to prevent division by zero.
In our paper, we use a modified version of this: we use $L(P_1,P_2) = \Xi(P_1, P_2) + \Xi(P_2, P_1) $ to make the divergence symmetric.
Besides the KL-divergence, other functions may be used as the loss function. For more details, see Appendix~\ref{sec:alt_loss}.

An optimization algorithm needs to be chosen for the training. In classical machine learning, the optimization algorithm is typically a variant of the gradient descent algorithm, $\vec{\theta}\leftarrow\vec{\theta}-\alpha \vec\nabla_{\vec{\theta}} L$, where the parameters $\vec\theta$ are updated along the direction where the loss function decreases the fastest. In principle, the same algorithm can be applied in quantum machine learning. However, in general, backpropagation cannot be performed for a quantum machine learning model. Alternative methods to compute the gradient are needed, such as the finite difference method or the parameter shift method, which are less efficient. Therefore, gradient-free optimization methods are often used, including the Powell method~\cite{powell_efficient_1964}, the constrained optimization by linear approximation (COBYLA) method~\cite{powell_direct_1998}, and the simultaneous perturbation stochastic approximation (SPSA) method~\cite{spall_overview_1998}. In this paper, the SPSA algorithm is used for optimization unless otherwise specified. For more details, see Appendix~\ref{sec:optimization}.

\section{Results \label{sec:results}}
To test the effectiveness of our approach, we trained our model to reconstruct certain quantum states. The results are presented in this section. To begin with, we tested our model on ideal simulators. As shown in Section~\ref{sec:results:ideal}, a range of different quantum states can be reconstructed.
Then, to verify compatibility with NISQ devices, we trained and executed the variational quantum circuit using IBM's and IonQ's quantum processors, successfully performing tomography for a 3-qubit Greenberger–Horne–Zeilinger state, as presented in Section~\ref{sec:results:qpu}.
In the aforementioned tests, the quantum states were measured in all Pauli bases; yet, the ability to perform tomography with fewer bases would be beneficial. To explore this possibility, using an ideal simulator, we performed tomography with a subset of the Pauli bases in Section~\ref{sec:results:incomplete}.

\subsection{Results with ideal simulator \label{sec:results:ideal} }
To test the performance of our method, we trained a variational quantum circuit to reconstruct a variety of pure quantum states, including the Greenberger–Horne–Zeilinger (GHZ) state, the ground state of a spin chain Hamiltonian, and several quantum states generated by random quantum circuits.
The details will be given below. 
Here, the target states are measured in all Pauli bases.

To start with, we tested our model on the GHZ state. An n-qubit GHZ state has the following form
\begin{equation}
    \ket{\text{GHZ}} = \frac{\ket{0}^{\otimes n}+\ket{1}^{\otimes n}}{\sqrt{2}}.
    \label{eqn:GHZ}
\end{equation}
It is the equal superposition of all spin-up state and all spin-down state and is sometimes referred to as the cat state. 
As an entangled state with non-local correlation, the GHZ state is an important state in quantum information. 
And it is often used as a benchmark for machine learning methods of quantum state tomography~\cite{li_experimental_2024,cha_attentionbased_2022,zhang_efficient_2022}. 
We also choose to use this state to benchmark our method.

To evaluate the performance of the model, we calculate the quantum fidelity between the target state and the reconstructed state at the end of training. Note, as mentioned before, experimentally, the quantum fidelity is not trivially accessible; here, the fidelity can be calculated, since the target state is known to us.
In our study, fidelity is not used during training. It is only used after training as a performance metric.

\begin{figure*}
    \centering
    \begin{subfigure}[c]{0.3\linewidth}
        \centering
        \includegraphics[width=\linewidth]{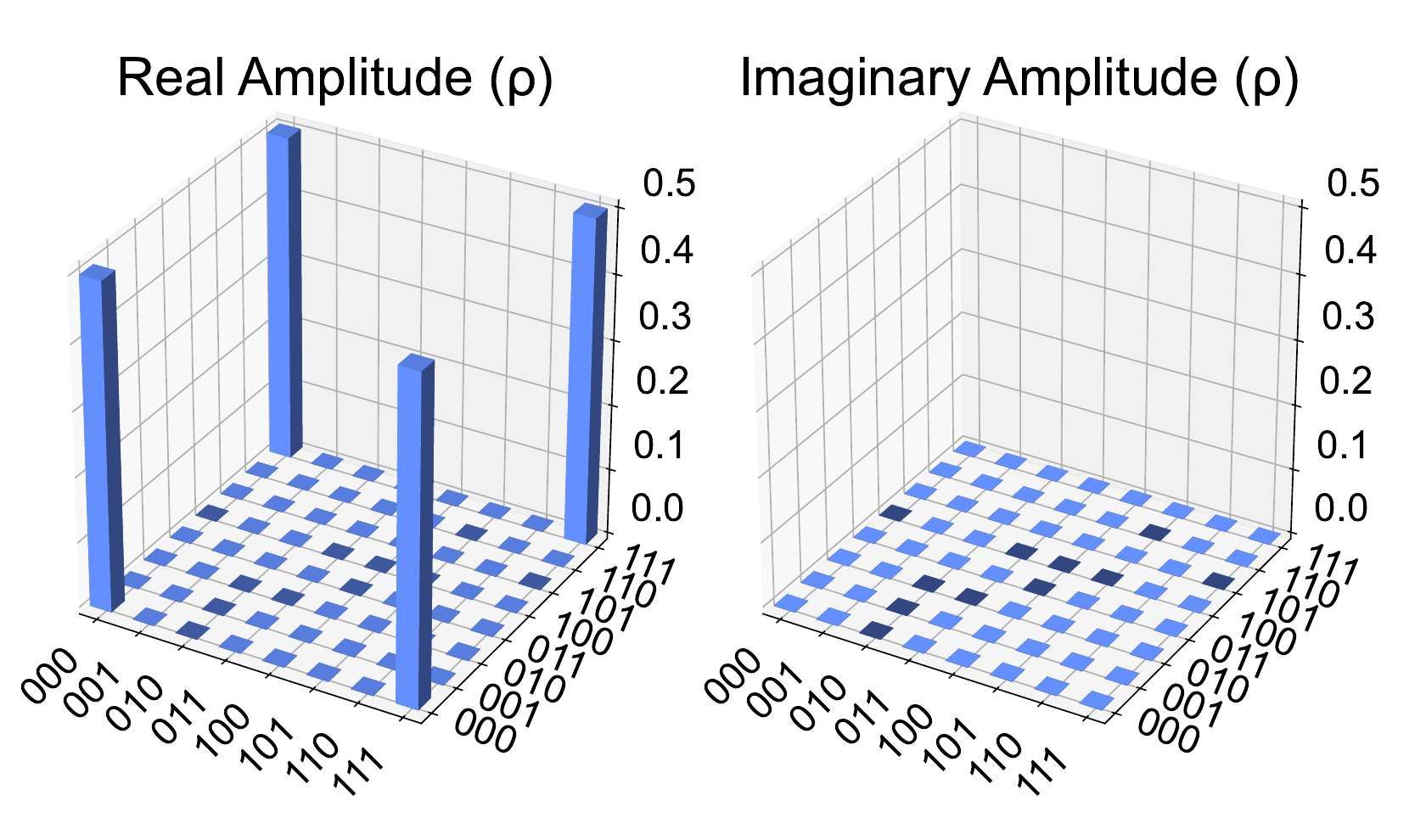}
        \caption{Ideal GHZ state}
        \label{fig:q3_ghz_state_city_spsa}
    \end{subfigure}
    \begin{subfigure}[c]{0.3\linewidth}
        \centering
        \includegraphics[width=\linewidth]{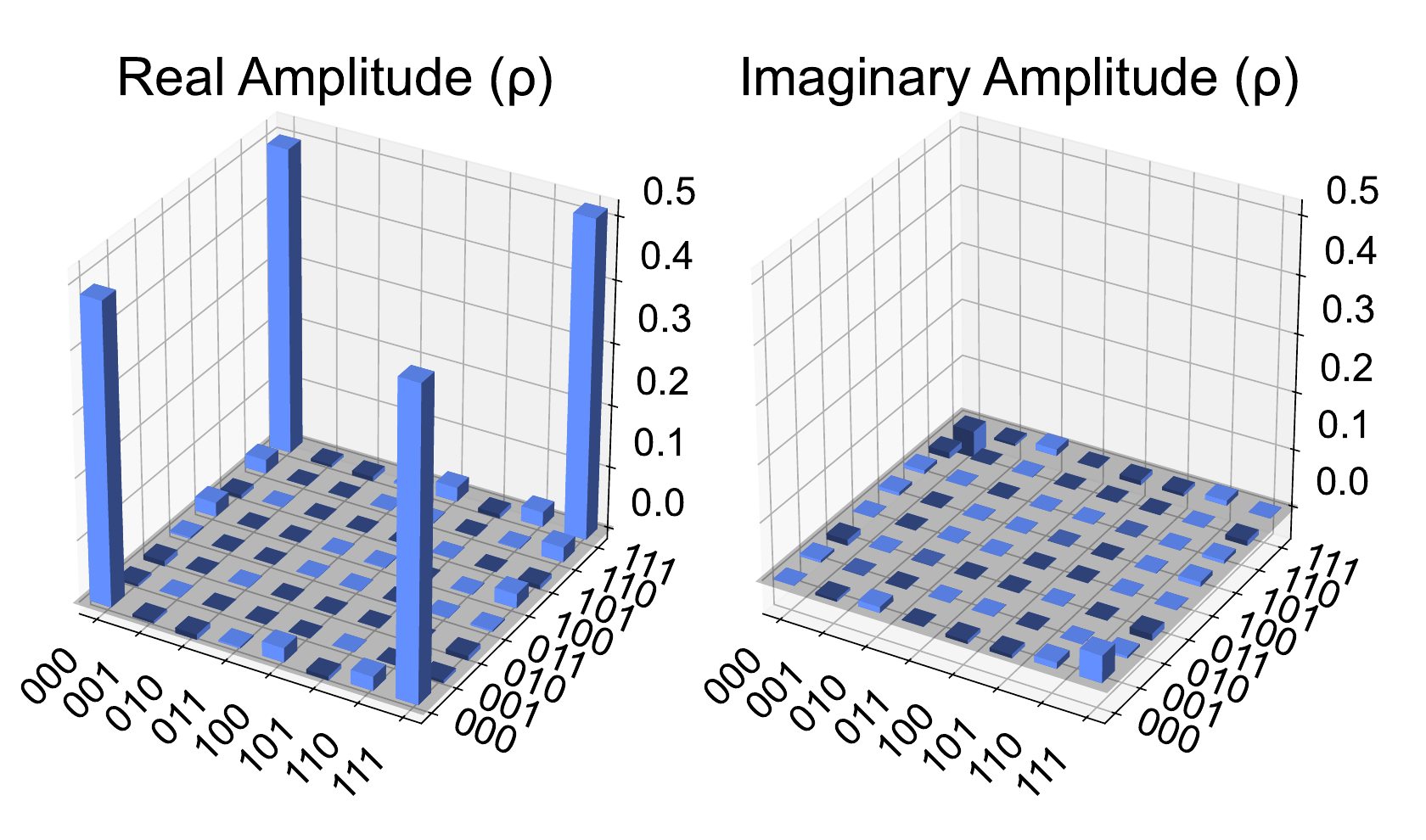}
        \caption{Reconstructed state}
        \label{fig:q3_reconstructed_state_city_spsa}
    \end{subfigure}
    \begin{subfigure}[c]{0.2\linewidth}
        \centering
        \includegraphics[width=\linewidth]{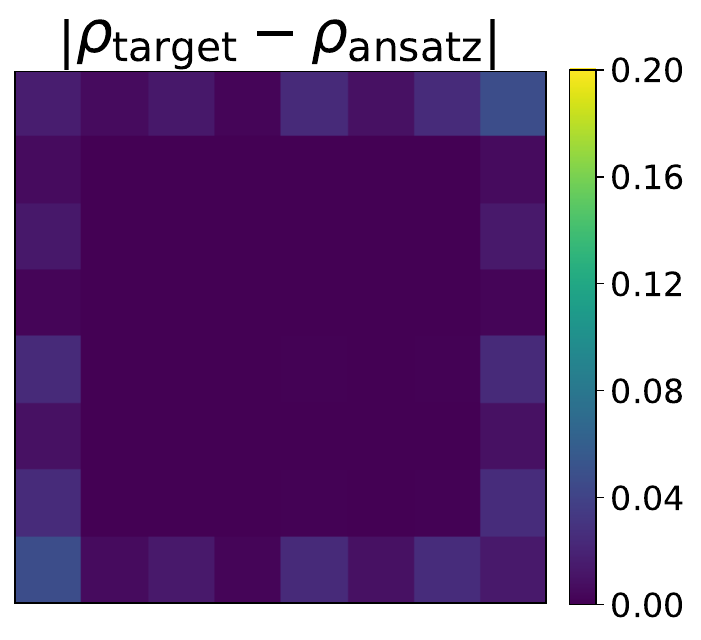}
        \caption{Absolute difference between (a) and (b)}
        \label{fig:dm_diff_q3_ghz}
    \end{subfigure}
    
    \caption{Reconstruction of the 3-qubit GHZ state. \ref{fig:q3_ghz_state_city_spsa} shows the visualization of the target state, while the \ref{fig:q3_reconstructed_state_city_spsa} shows the reconstructed state in a typical trial. The difference between them are calculated $\rho_\text{target}-\rho_\text{ansatz}$, and \ref{fig:dm_diff_q3_ghz} visualizes the magnitude of its matrix elements.
    }
    \label{fig:q3_ghz_state_visual_spsa}
\end{figure*}

\begin{figure*}
    \centering
    \begin{subfigure}[c]{0.3\linewidth}
        \centering
        \includegraphics[width=\linewidth]{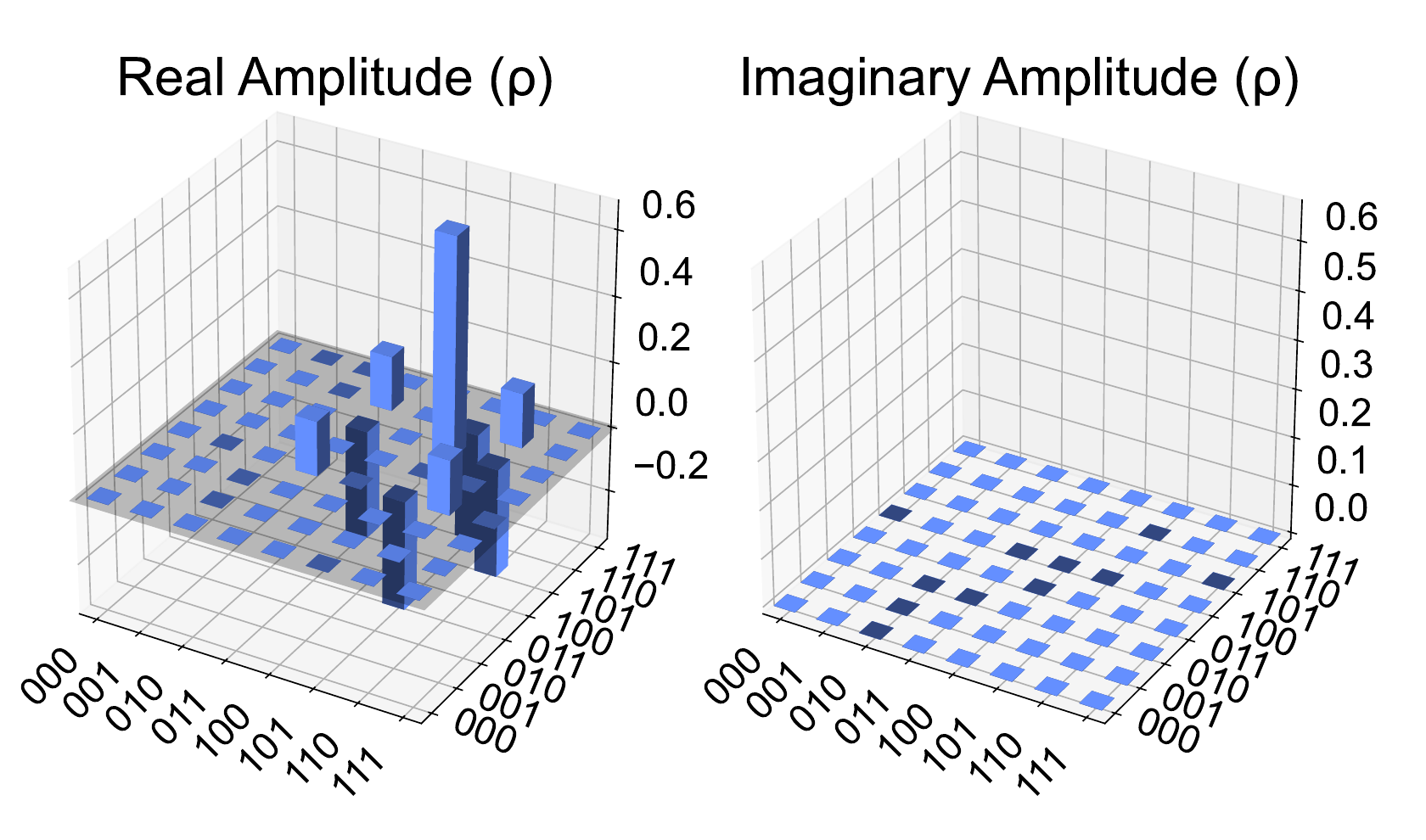}
        \caption{Spin chain ground state}
        \label{fig:q3_spch_state_city_spsa}
    \end{subfigure}
    \begin{subfigure}[c]{0.3\linewidth}
        \centering
        \includegraphics[width=\linewidth]{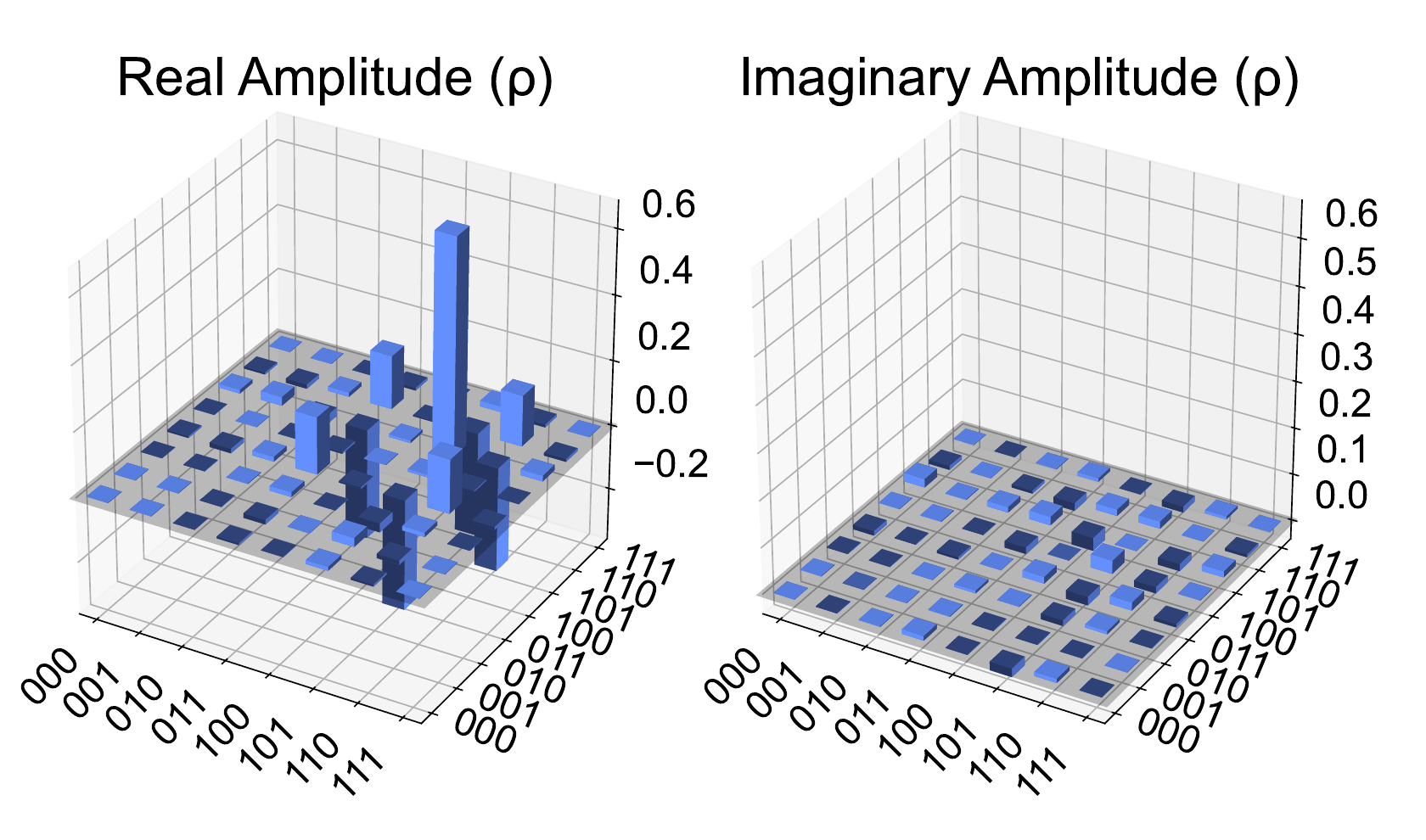}
        \caption{Reconstructed state}
        \label{fig:q3_spch_reconstructed_state_city_spsa}
    \end{subfigure}
    \begin{subfigure}[c]{0.2\linewidth}
        \centering
        \includegraphics[width=\linewidth]{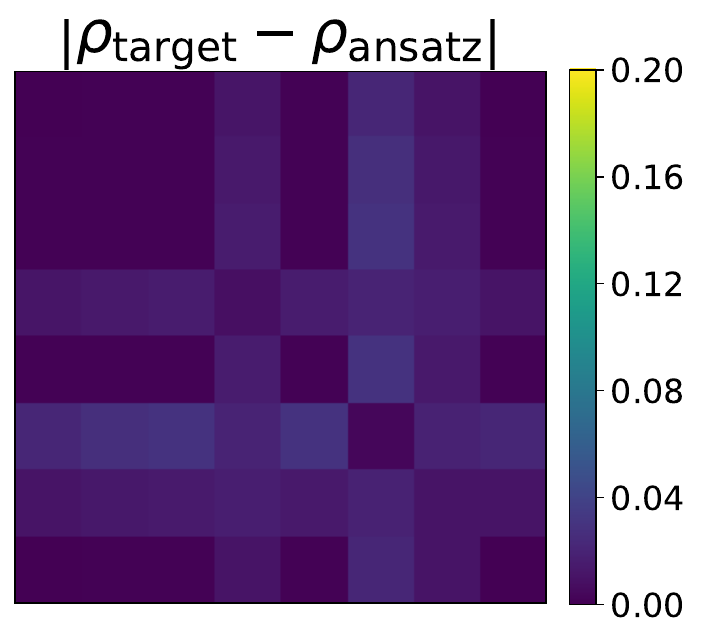}
        \caption{Absolute difference between (a) and (b).}
        \label{fig:dm_diff_q3_spch}
    \end{subfigure}
    
    \caption{Reconstruction of the 3-qubit spin chain ground state. \ref{fig:q3_spch_state_city_spsa} shows the visualization of the target state, while the \ref{fig:q3_spch_reconstructed_state_city_spsa} shows the reconstructed state in a typical trial.  The difference between them are calculated $\rho_\text{target}-\rho_\text{ansatz}$, and \ref{fig:dm_diff_q3_spch} visualizes the magnitude of its matrix elements.
    }
    \label{fig:q3_spch_state_visual_spsa}
\end{figure*}

\begin{figure}
    \centering
    \begin{subfigure}[t]{0.46\linewidth}
        \centering
        \includegraphics[width=\linewidth]{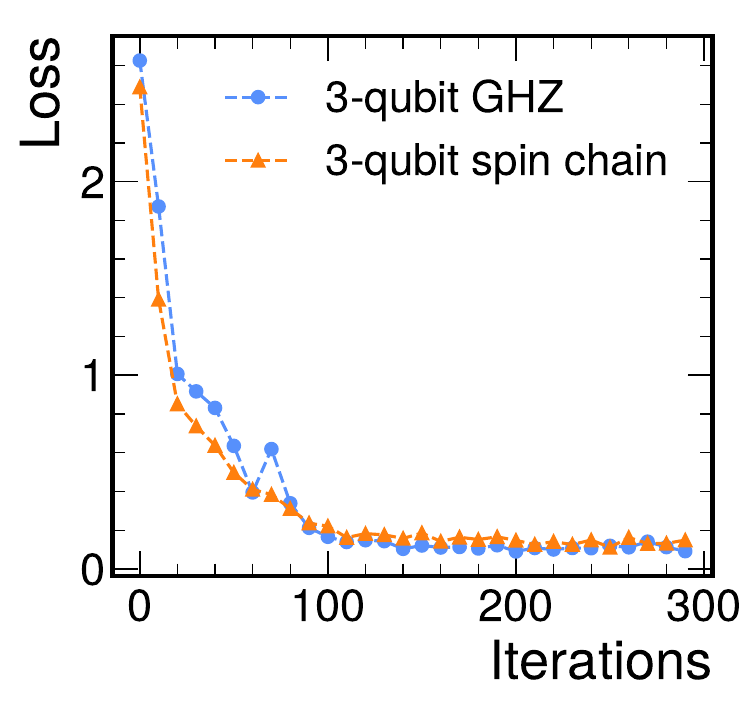}
        \caption{3 qubits}
        \label{fig:loss_vs_iterations_summary:q3}
    \end{subfigure}
    \begin{subfigure}[t]{0.46\linewidth}
        \centering
        \includegraphics[width=\linewidth]{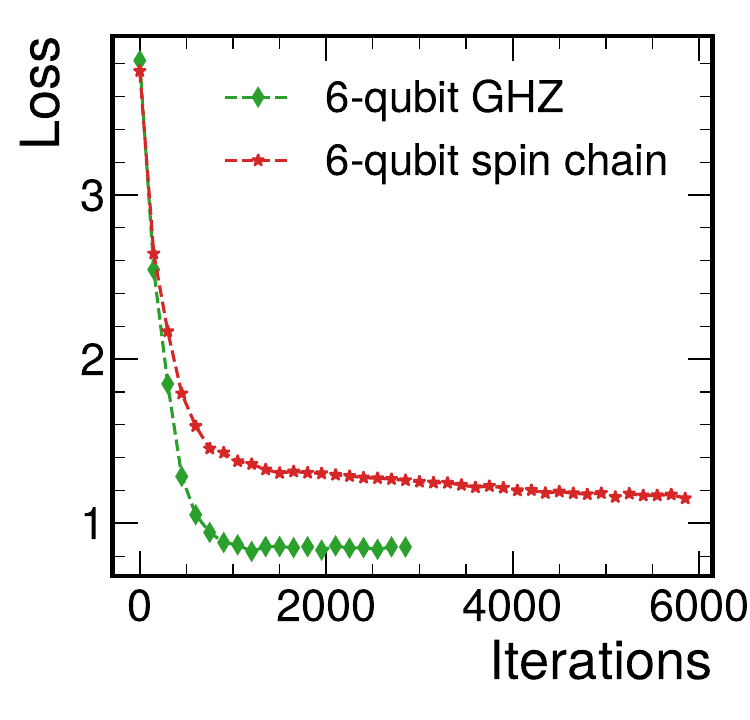}
        \caption{6 qubits}
        \label{fig:loss_vs_iterations_summary:q6}
    \end{subfigure}
    
    \caption{Loss function during training. The loss function is plotted against training iterations in a typical trial for the 3-qubit and 6-qubit GHZ states and spin chain ground states.}
    \label{fig:loss_vs_iterations_summary}
\end{figure}

\begin{figure*}[t]
    \centering
    \begin{subfigure}[t]{0.24\linewidth}
        \centering
        \includegraphics[width=\linewidth]{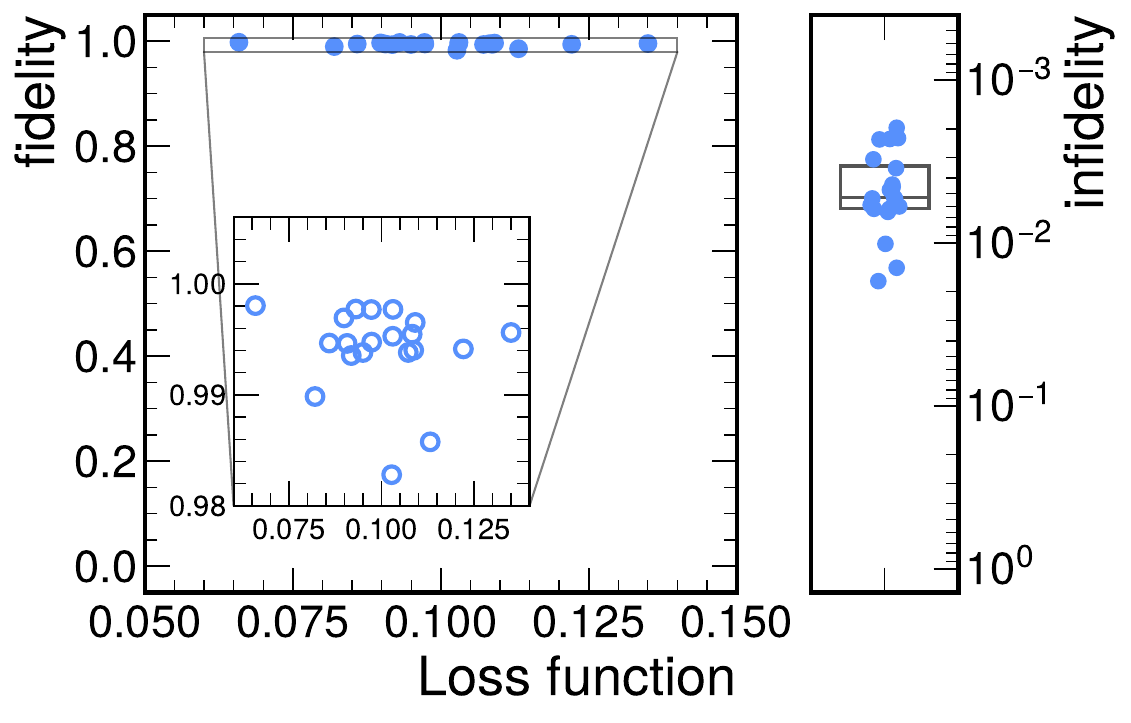}
        \caption{3-qubit GHZ state}
        \label{fig:panel:q3_ghz}
    \end{subfigure}
    \hfill
    \begin{subfigure}[t]{0.24\linewidth}
        \centering
        \includegraphics[width=\linewidth]{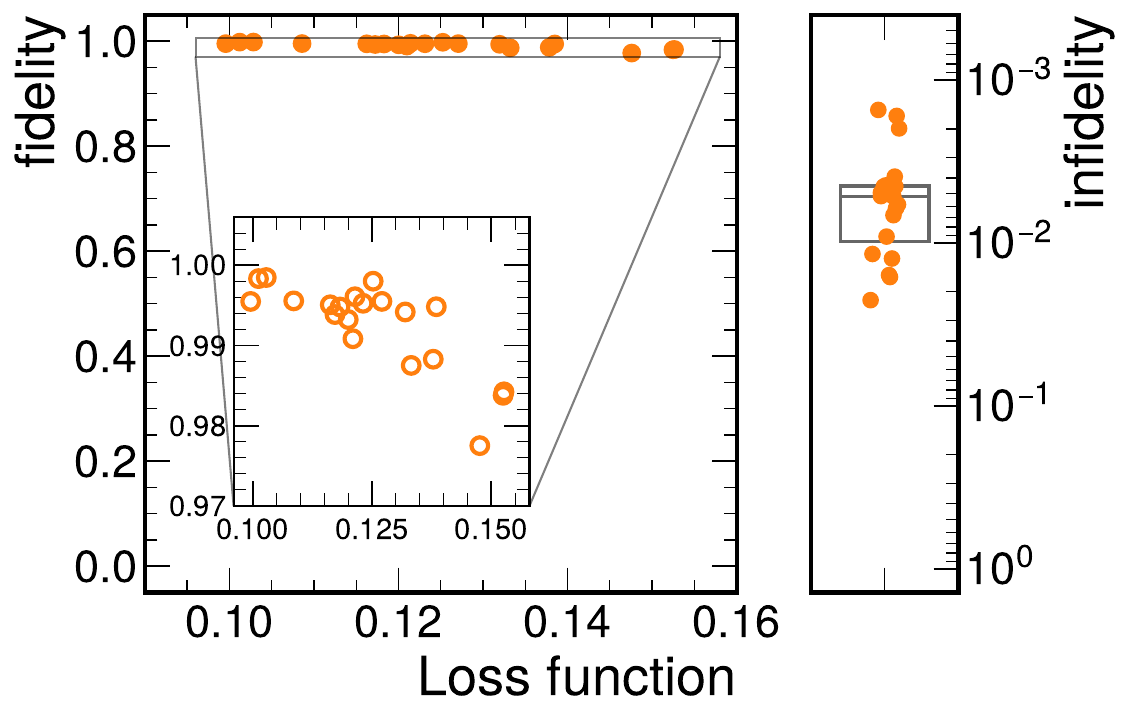}
        \captionsetup{justification=centering}
        \caption{3-qubit spin chain ground state}
        \label{fig:panel:q3_spch}
    \end{subfigure}
    \hfill
    \begin{subfigure}[t]{0.24\linewidth}
        \centering
        \includegraphics[width=\linewidth]{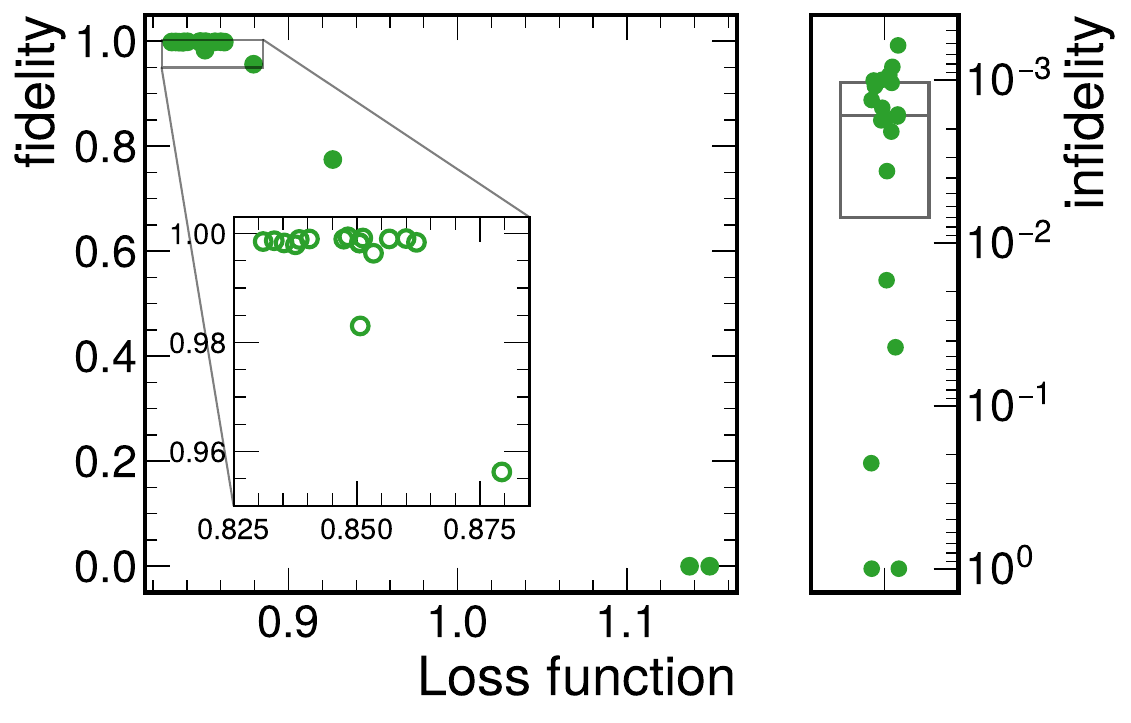}
        \caption{6-qubit GHZ state}
        \label{fig:panel:q6_ghz}
    \end{subfigure}
    \hfill
    \begin{subfigure}[t]{0.24\linewidth}
        \centering
        \includegraphics[width=\linewidth]{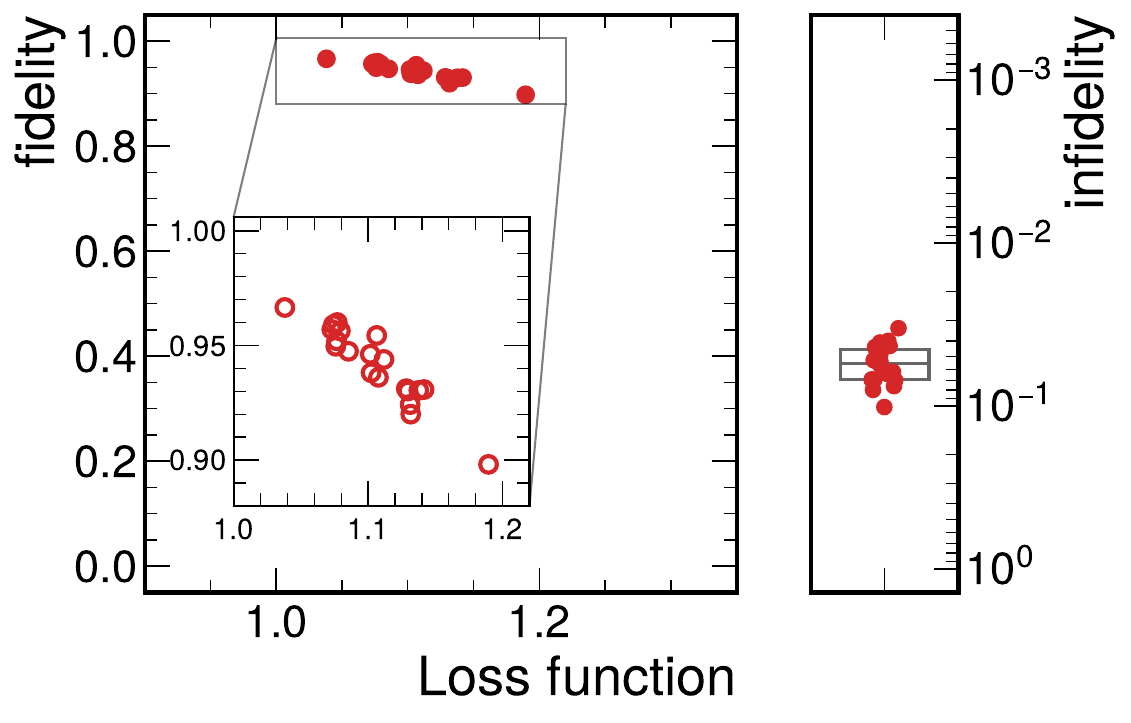}
        \captionsetup{justification=centering}
        \caption{6-qubit spin chain ground state}
        \label{fig:panel:q6_spch}
    \end{subfigure}

    \caption{Reconstruction of the 3-qubit and 6-qubit GHZ states and spin chain ground states. In each case, the left figure plots the joint distribution of fidelity and final loss in all 20 trials. The right figure plots the distribution of infidelity, defined as $1$ $-$ fidelity. }
    \label{fig:panel}
\end{figure*}

\newcommand{\scaleA}{0.22}
\begin{figure*}[bt]
    \centering
    \begin{subfigure}[t]{\scaleA \linewidth}
        \centering
        \includegraphics[width=\linewidth]{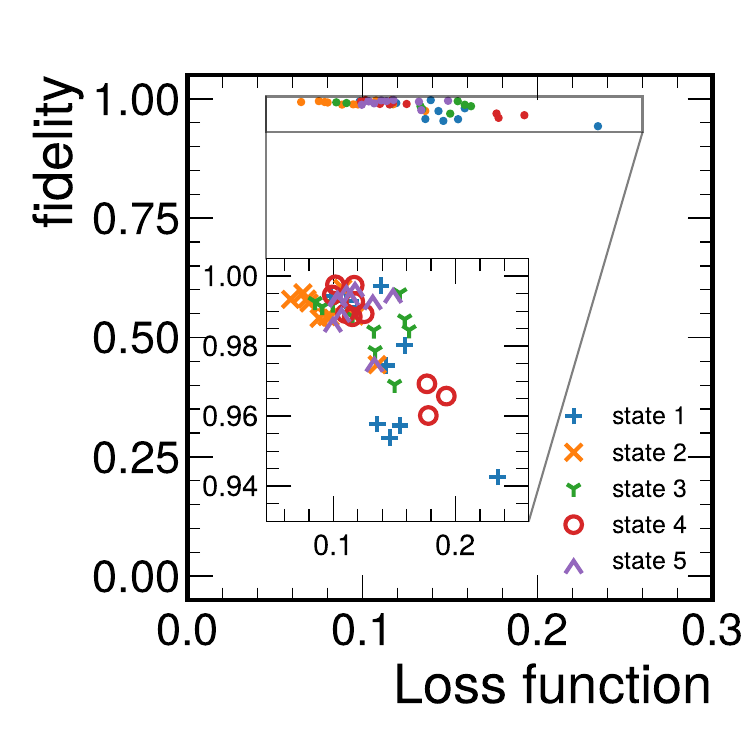}
        \captionsetup{justification=centering}
        \caption{Fidelity and loss 3-qubit~case}
        \label{fig:random_states_q3:fidelity_loss}
    \end{subfigure}
    \hfill
    \begin{subfigure}[t]{\scaleA\linewidth}
        \centering
        \includegraphics[width=\linewidth]{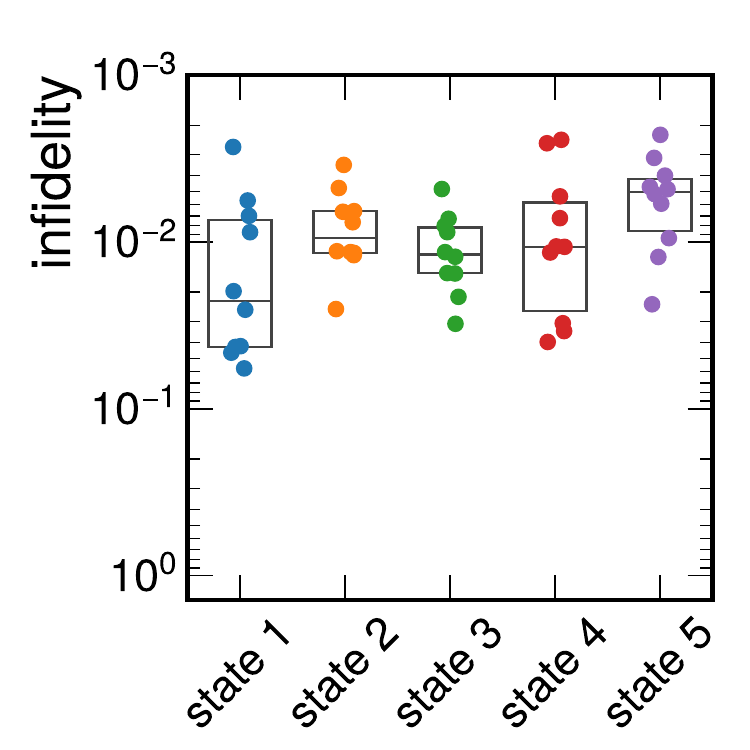}
        \captionsetup{justification=centering}
        \caption{Infidelity \newline 3-qubit~case }
        \label{fig:random_states_q3:fidelity_stripplot}
    \end{subfigure}
    \hfill
    \begin{subfigure}[t]{\scaleA\linewidth}
        \centering
        \includegraphics[width=\linewidth]{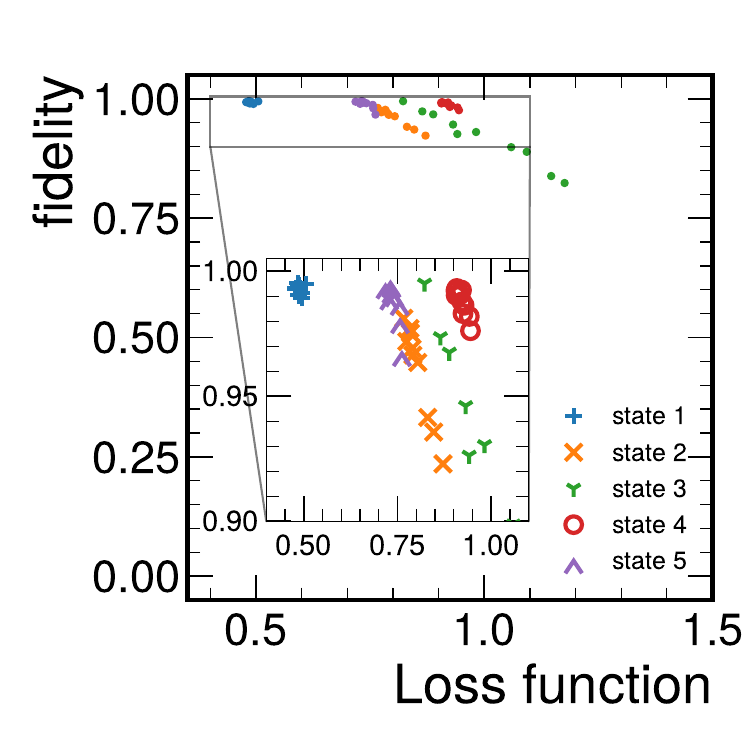}
        \captionsetup{justification=centering}
        \caption{Fidelity and loss   6-qubit~case}
        \label{fig:random_states_q6:fidelity_loss}
    \end{subfigure}
    \hfill
    \begin{subfigure}[t]{\scaleA\linewidth}
        \centering
        \includegraphics[width=\linewidth]{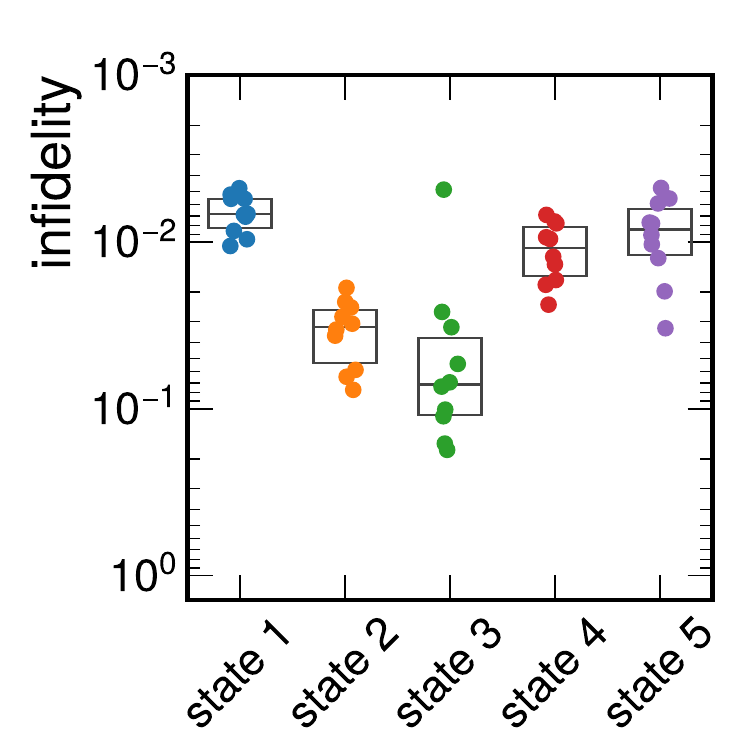}
        \captionsetup{justification=centering}
        \caption{Infidelity \newline  6-qubit~case}
        \label{fig:random_states_q6:fidelity_stripplot}
    \end{subfigure}
    \caption{Reconstruction of random circuit states. \ref{fig:random_states_q3:fidelity_loss} shows the joint distribution of fidelity and loss function for the five state generated from random 3-qubit circuits, and \ref{fig:random_states_q3:fidelity_stripplot} visualizes the distribution of infidelity for each state. Likewise, \ref{fig:random_states_q6:fidelity_loss} and \ref{fig:random_states_q6:fidelity_stripplot} present the results for 6-qubit random circuit states.}
    \label{fig:random_states}
\end{figure*}

\subparagraph{3-qubit GHZ state}
As a proof of concept, we tried quantum state tomography for a 3-qubit GHZ state. The GHZ state was measured in all 27 bases for 100 shots each. We used the ansatz in Fig.~\ref{fig:ansatz} with 10 layers. 
The $\epsilon$ in Equation \ref{eqn:kl} was set to $10^{-3}$.
The training was performed with the SPSA optimization algorithm. 
To visualize the result, we plot the density matrix of the target GHZ state and the reconstructed state in Fig.~\ref{fig:q3_ghz_state_visual_spsa}; the values of the loss function during training are plotted in Fig.~\ref{fig:loss_vs_iterations_summary}. This illustrates that the variational quantum circuit can reconstruct the GHZ state.
Furthermore, to test the stability of this procedure, we trained the model 20 times and collected the results. 
The median fidelity is about $0.995$.
The results are shown in Fig.~\ref{fig:panel:q3_ghz}. The distribution of obtained fidelity is plotted against the final loss, i.e., the loss function after the training completes. Additionally, the distribution of infidelity is plotted, which is defined as $1$ $-$ fidelity. As shown in the figure, the infidelity is within $0.01$ most of the time.

\subparagraph{6-qubit GHZ state}
To test the scalability of our method, we used the model to reconstruct the 6-qubit GHZ state. Here, all 729 bases were measured. For each basis, 100 measurements were taken. The ansatz in Fig.~\ref{fig:ansatz} with 6 qubits and 10 layers was used.
Training was performed using the SPSA algorithm. 
In Fig.~\ref{fig:loss_vs_iterations_summary}, we plot the loss function during training over the number of iterations in a typical run.
We ran the model 20 times. The performance of the model in these 20 runs is shown in Fig.~\ref{fig:panel:q6_ghz}.
In most trials, high fidelity is obtained. There are 3 outliers where the fidelity is significantly lower than in other trials. 
When compared to other points, these outliers can be identified using the loss function, since the final losses of these three are higher than the other ones.
In most trials, the fidelity is close to $0.998$. The median of the 20 runs is $0.998$.

\bigskip

The above results illustrate the feasibility of reconstructing the GHZ state. 
In the mean time, it is also important to show that this procedure can work for more general states.
Hence, we tested the ground state of a spin chain Hamiltonian. The results will reflect the effectiveness of the methodology on quantum states found in physical systems.
Following \cite{liu_variational_2020}, we choose the XXZ Heisenberg model, which has the Hamiltonian 
\begin{equation}
    H_{XXZ}=\sum_{l=1}^{L-1}\left[ J\left(\sigma_l^x \sigma_{l+1}^x + \sigma_l^y \sigma_{l+1}^y\right) + \Delta \sigma_l^z \sigma_{l+1}^z\right] + h \sum_{l=1}^{L} \sigma_l^z.
    \label{eqn:XXZ}
\end{equation}
We take measurements on the ground state of this Hamiltonian, which can be calculated using exact diagonalization for small systems. Here, we set the parameter $J$, $h$, and $\Delta$ to $1$.

\subparagraph{3-qubit spin chain ground state}
We used our model to perform state reconstruction for the ground state of the spin chain Hamiltonian in Equation~\ref{eqn:XXZ} for the 3-qubit case. Similar to the 3-qubit GHZ case, a 10-layer ansatz was used. Other training settings were also similar.
The spin chain ground state is visualized in Fig.~\ref{fig:q3_spch_state_visual_spsa}. The training loss of a typical trial is plotted in Fig.~\ref{fig:loss_vs_iterations_summary}.
The training results are shown in Fig.~\ref{fig:panel:q3_spch}. The performance is similar to the 3-qubit GHZ case. In most trials, the infidelity is below $0.01$. 
The median fidelity of 20 trials is $0.995$.

\subparagraph{6-qubit spin chain ground state}
We reconstructed the ground state of a 6-qubit spin chain. For each basis, 100 measurements were taken. A 16-layer version of the ansatz in Fig.~\ref{fig:ansatz} was used. Other than that, training settings were similar to the 6-qubit GHZ state. The result is shown in Fig.~\ref{fig:panel:q6_spch}. The obtained quantum fidelity is not as high as in the 6-qubit GHZ state case. Nevertheless, fidelity around or above 0.9 is obtained. 
The median fidelity of 20 trials is near $0.95$.

\begin{figure*}
    \centering
    \begin{subfigure}[t]{0.24\linewidth}
        \centering
        \includegraphics[width=\linewidth]{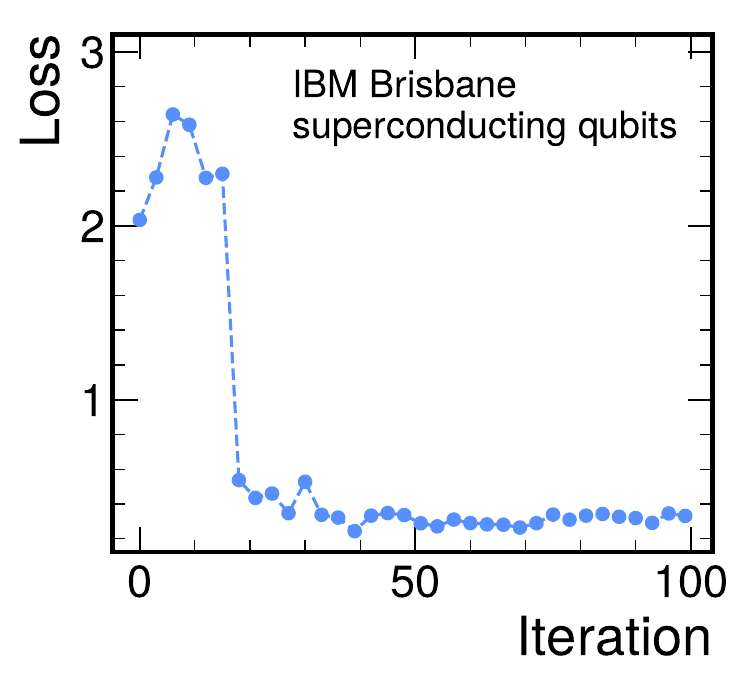}
        \caption{IBM Brisbane.}
        \label{fig:loss_ghz_ibm}
    \end{subfigure}
    \begin{subfigure}[t]{0.4\linewidth}
        \centering
        \includegraphics[width=\linewidth]{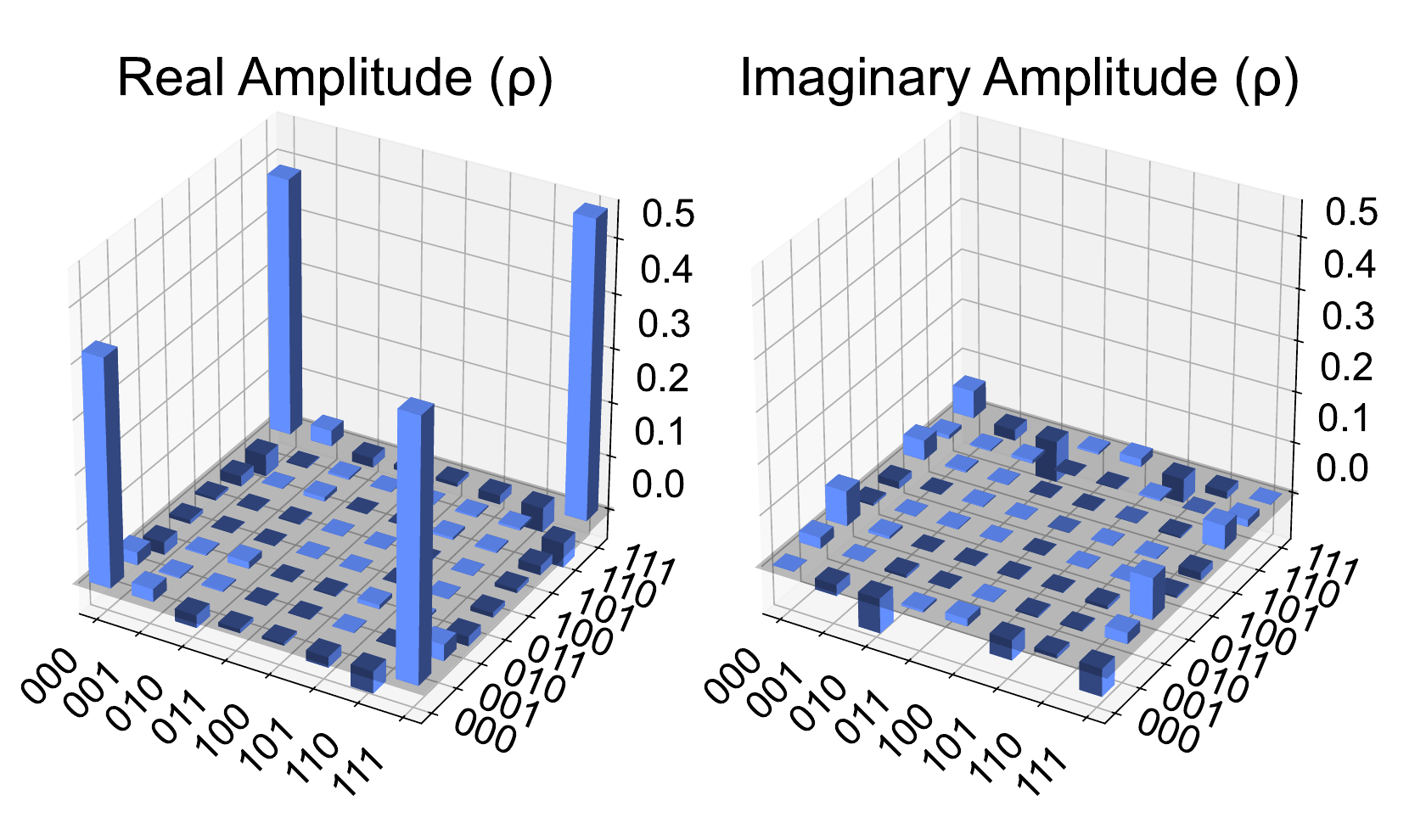}
        \caption{Reconstructed state (IBM)}
        \label{fig:q3_ghz_state_city_ibm}
    \end{subfigure}
    \begin{subfigure}[t]{0.24\linewidth}
        \centering
        \includegraphics[width=\linewidth]{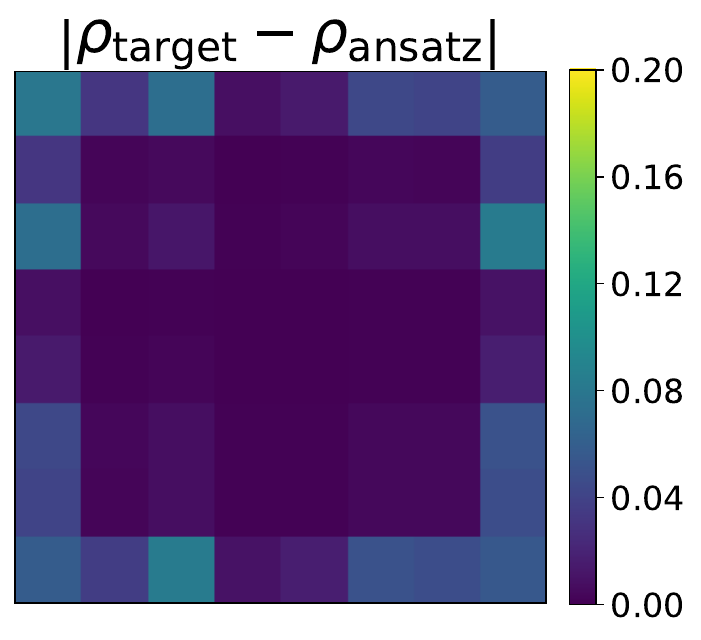}
        \caption{Reconstruction error (IBM)}
        \label{fig:dm_diff_q3_ghz_ibm}
    \end{subfigure}

    \begin{subfigure}[t]{0.24\linewidth}
        \centering
        \includegraphics[width=\linewidth]{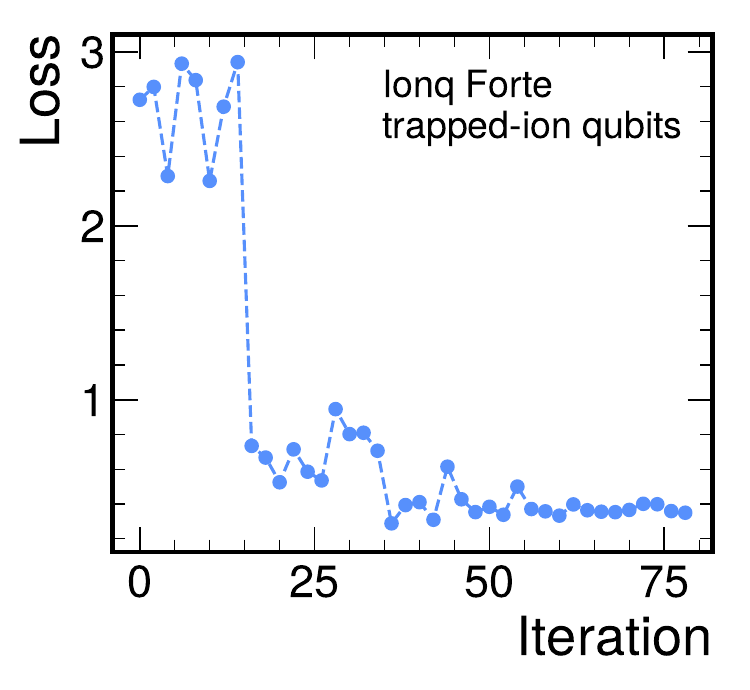}
        \caption{IonQ Forte.}
        \label{fig:loss_ghz_ionq}
    \end{subfigure}
    \begin{subfigure}[t]{0.4\linewidth}
        \centering
        \includegraphics[width=\linewidth]{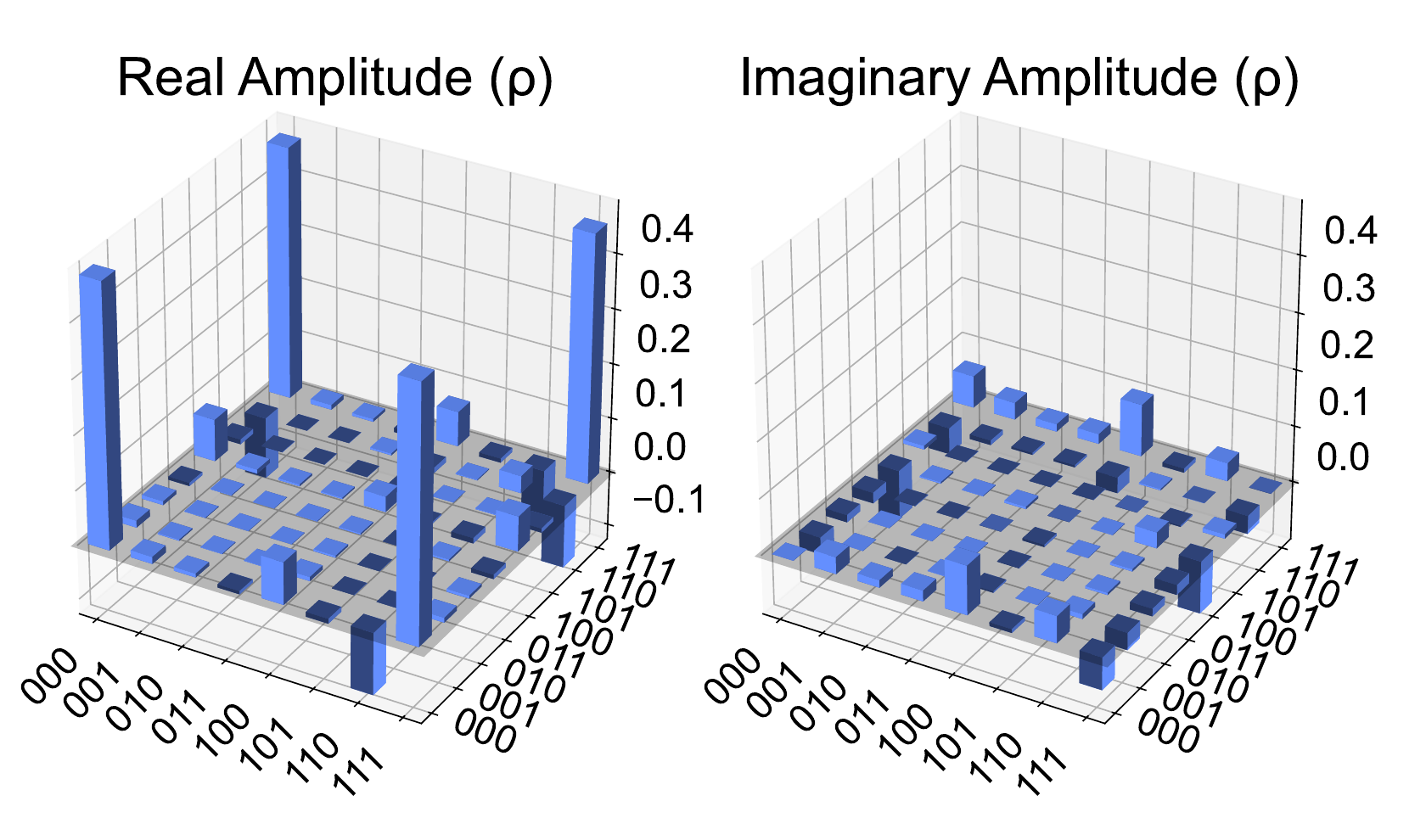}
        \caption{Reconstructed state (IonQ)}
        \label{fig:q3_reconstructed_state_city_ionq}
    \end{subfigure}
    \begin{subfigure}[t]{0.24\linewidth}
        \centering
        \includegraphics[width=\linewidth]{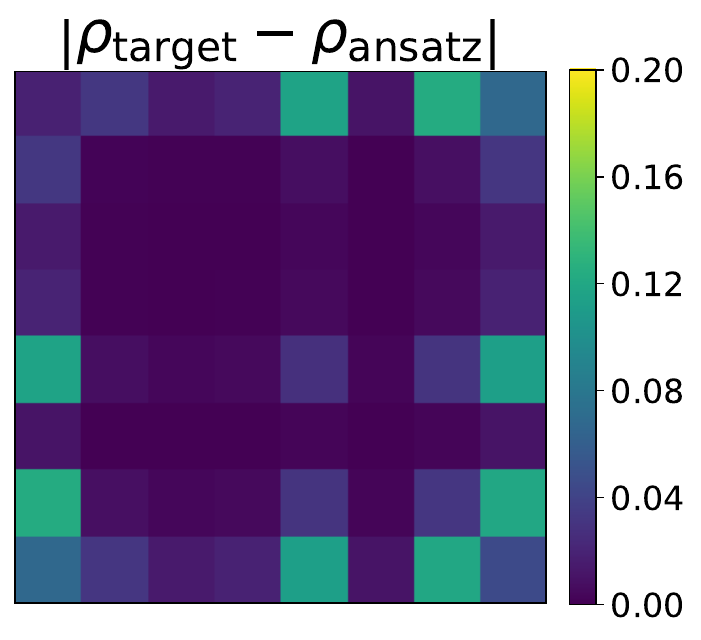}
        \caption{Reconstruction error (IonQ)}
        \label{fig:dm_diff_q3_ionq}
    \end{subfigure}

    \caption{3-qubit GHZ state reconstruction with quantum computers. A variational circuit was trained on IBM Brisbane. \ref{fig:loss_ghz_ibm} plots the training loss, \ref{fig:q3_ghz_state_city_ibm} visualizes the reconstructed state at the end of training, and \ref{fig:dm_diff_q3_ghz_ibm} shows the deviation of reconstructed state from a GHZ state. Likewise, training was performed on IonQ Forte. The results are similarly presented in \ref{fig:loss_ghz_ionq}, \ref{fig:q3_reconstructed_state_city_ionq}, and \ref{fig:dm_diff_q3_ionq}. }
    \label{fig:qpu}
    
\end{figure*}

\smallskip

\subparagraph{Quantum states from random circuits} 
To further illustrate the effectiveness of our approach, we obtained random circuits using Qiskit \verb|random_circuit| function and performed tomography on the states thus generated. We created random 3-qubit circuits with 5 layers of single-qubit or two-qubit gates.
The circuits are shown in Appendix~\ref{sec:random_circuits}.
Using this method, we obtained five different states. Then, we attempted to use a 10-layer variational quantum circuit to reconstruct them from the measurement data. In each of the 27 bases, 100 measurements were performed. For each state, the training was repeated 10 times. The results are presented in Figs.~\ref{fig:random_states_q3:fidelity_loss} and \ref{fig:random_states_q3:fidelity_stripplot}. As shown in the figure, fidelity over $0.9$ is consistently obtained. In many trials, fidelity near or above $0.98$ is obtained.
Similarly, we created five different 6-qubit circuits and performed tomography on them with a 12-layer ansatz. As presented in Figs.~\ref{fig:random_states_q6:fidelity_loss} and \ref{fig:random_states_q6:fidelity_stripplot}, the performance varies between different states, but fidelity over $0.9$ can be typically obtained.

\begin{figure*}
    \centering
    \vspace{1em}
    \begin{subfigure}[t]{0.24\linewidth}
        \centering
        \includegraphics[width=\linewidth]{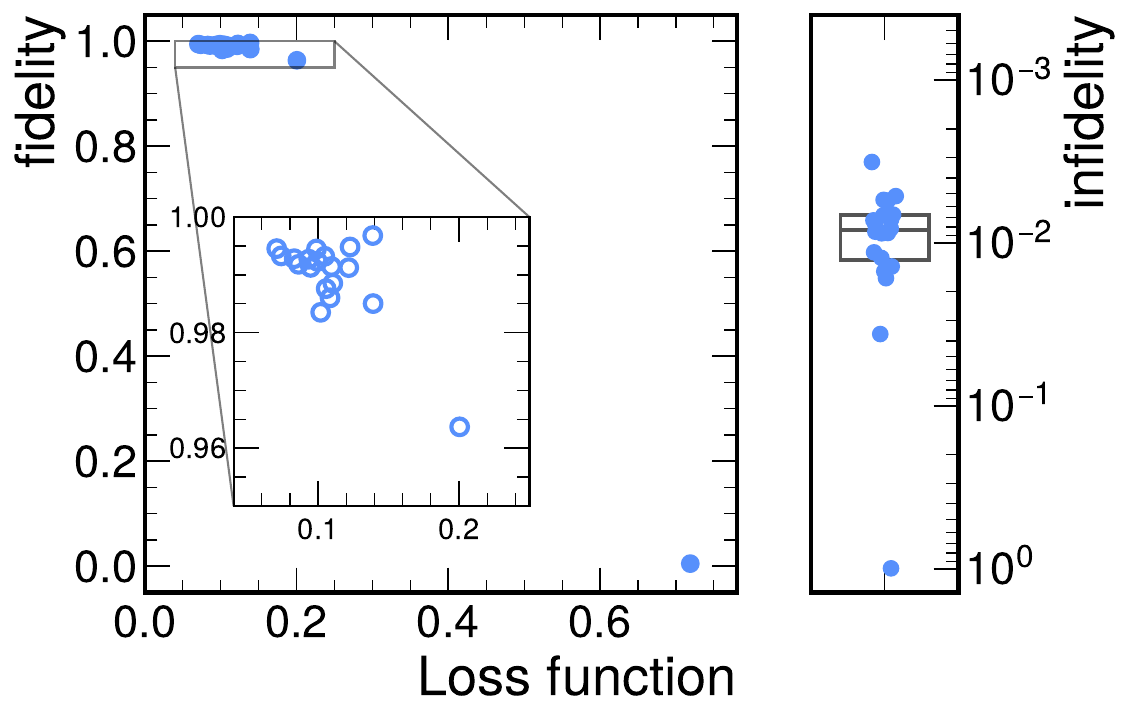}
        \captionsetup{justification=centering}
        \caption{3-qubit GHZ with 15 bases}
        \label{fig:panel_ic:q3_ghz_b15}
    \end{subfigure}
    \hspace{0.2em}
    \begin{subfigure}[t]{0.24\linewidth}
        \centering
        \includegraphics[width=\linewidth]{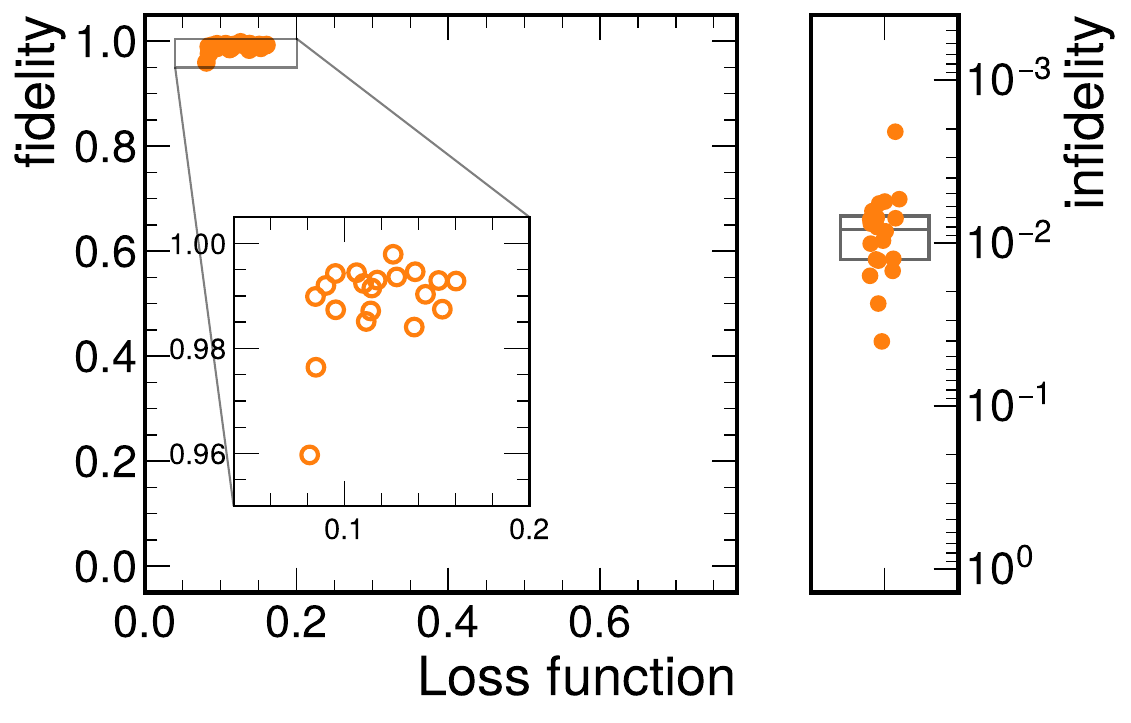}
        \captionsetup{justification=centering}
        \caption{3-qubit spin chain ground state with 15 bases}
        \label{fig:panel_ic:q3_spch_b15}
    \end{subfigure}    
    \begin{subfigure}[t]{0.24\linewidth}
        \centering
        \includegraphics[width=\linewidth]{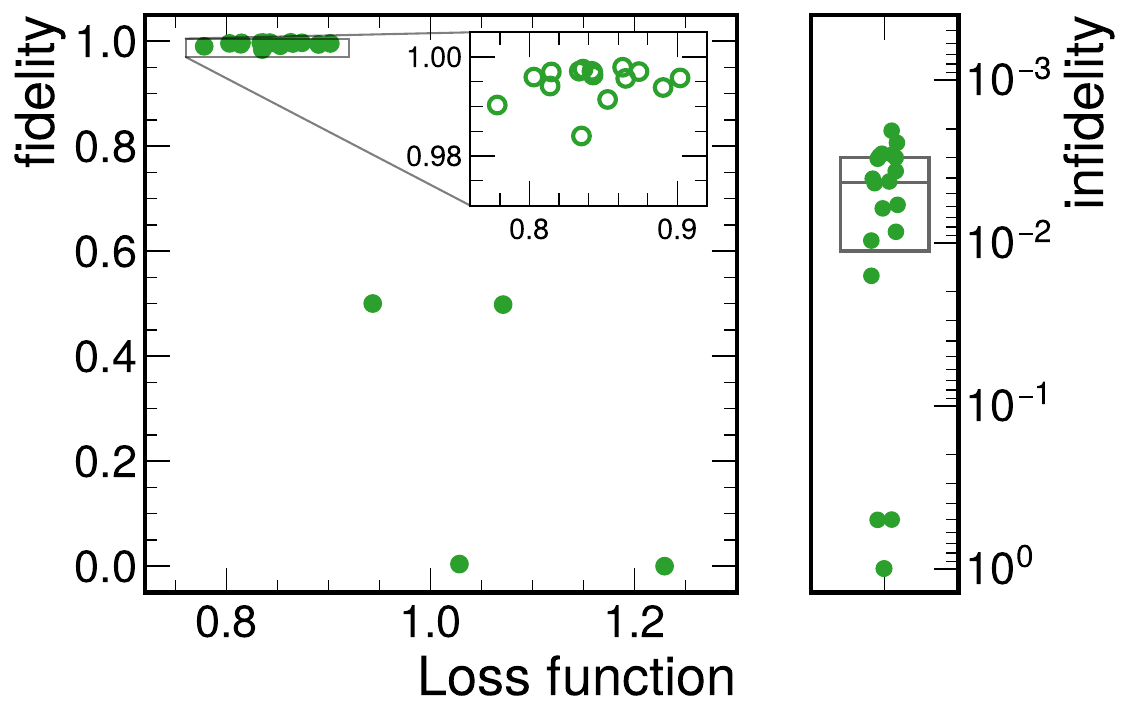}
        \captionsetup{justification=centering}
        \caption{6-qubit GHZ with 200 bases}
        \label{fig:panel_ic:q6_ghz_b200}
    \end{subfigure}
    \hspace{0.2em}
    \begin{subfigure}[t]{0.24\linewidth}
        \centering
        \includegraphics[width=\linewidth]{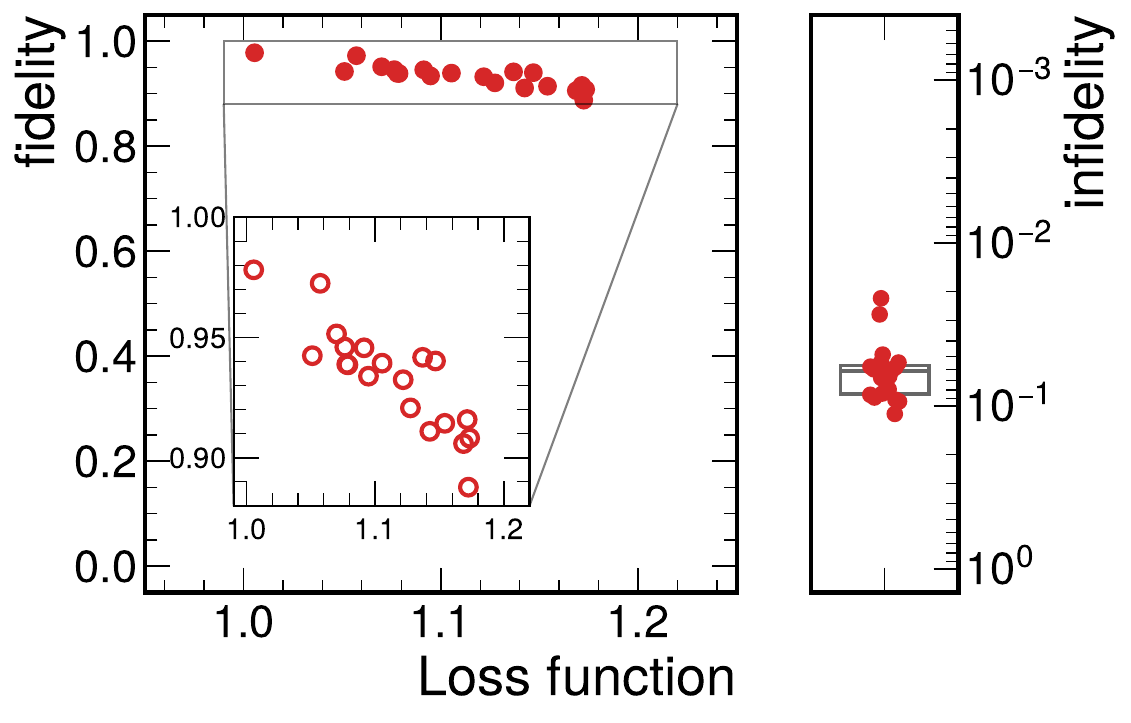}
        \captionsetup{justification=centering}
        \caption{6-qubit spin chain ground state with 200 bases}
        \label{fig:panel_ic:q6_spch_b200}
    \end{subfigure}

    \caption{QST with incomplete measurement bases. Rather than all Pauli bases, only a subset was measured. In the 3-qubit case, 15 bases were randomly chosen. The results are shown in \ref{fig:panel_ic:q3_ghz_b15} for GHZ state and \ref{fig:panel_ic:q3_spch_b15} for spin chain ground state. In the 6-qubit case, 200 bases were randomly chosen. The results are shown in \ref{fig:panel_ic:q6_ghz_b200} for GHZ state and \ref{fig:panel_ic:q6_spch_b200} for spin chain ground state.}
    \label{fig:panel_ic}
\end{figure*}

\subsection{Results with quantum hardware \label{sec:results:qpu}}
The above results illustrate the feasibility of our methodology on ideal simulator. In this section, we take a step further by training and executing the QML model on quantum hardware.
We use the 3-qubit GHZ state as the target state. 
The measurement data for the target state was produced by a classical simulator, while the variational quantum circuit was trained and executed on quantum hardware.
To save computation resource, a 5-layer ansatz was used. 
The COBYLA optimization algorithm~\cite{powell_direct_1998} was used.
After the training finished, we calculated the quantum fidelity to evaluate the performance.
The variational quantum circuit was trained on IBM Brisbane, and a fidelity of 0.968 was obtained at the end. 
With similar settings, the model was also trained on IonQ's Forte and achieved a fidelity of 0.933. The loss function during training and a visualization of the reconstructed state are presented in Fig.~\ref{fig:qpu} for both IBM and IonQ runs.

\subsection{Results with incomplete bases \label{sec:results:incomplete}}

In the above tests, we took measurements in all measurement bases. For a 3-qubit state, the total number of bases is $3^3=27$, whereas for a 6-qubit state, the total number of bases is $3^6=729$. 
Measuring all bases is required to pinpoint a generic quantum state; however, under suitable assumptions, a smaller number of bases may suffice~\cite{gross_quantum_2010}. In our case, our goal is to reconstruct states which can be produced by variational circuits within a reasonable depth and can be described by a polynomial number of parameters. Hence, it is likely that tomography can be performed with a reduced set of measurement bases. The ability to do this would reduce the required resource for QST and is a prerequisite for tackling the exponential scaling problem in QST. To verify whether this is doable, we conducted QST for the GHZ state and the spin chain ground state with incomplete measurements. As in Section~\ref{sec:results:ideal}, the training was performed on an ideal simulator.

\subparagraph{3-qubit GHZ state with 15 bases}
Here, we took measurements from 15 different randomly chosen bases. We repeated the trials for 20 times. The measurement bases for different trials were chosen independently. For each basis, 100 measurements were taken. Training was performed with the SPSA algorithm. The result is shown in Fig.~\ref{fig:panel_ic:q3_ghz_b15}. High fidelity is obtained except for one trial. For most trials, the fidelity is above 0.98. The median fidelity is $0.992$.

\subparagraph{3-qubit spin chain ground state with 15 bases}
Likewise, we performed QST on the 3-qubit spin chain ground state with 15 measurement bases. The training settings were similar. Again, in most trials, fidelity over 0.98 can be obtained, with the median being $0.992$. The results are shown in Fig.~\ref{fig:panel_ic:q3_spch_b15}.

\subparagraph{6-qubit GHZ state with 200 bases}
Similarly, we attempted tomography with incomplete bases for the 6-qubit GHZ state. We took 200 random bases out of all 729 for each trial. For each basis, 100 measurements were taken. The results are shown in Fig.~\ref{fig:panel_ic:q6_ghz_b200}. 
There are 4 outliers with visibly lower fidelity.
However, most of the time, fidelity near 0.99 or above is achieved.
The median of the 20 runs is $0.996$

\subparagraph{6-qubit spin chain ground state with 200 bases}
Under similar settings, tomography was performed for the ground state of the 6-qubit spin chain Hamiltonian. 200 random bases were measured for each trial. Fidelity over 0.9 was obtained except for one trial at 0.89. The median fidelity in 20 runs is $0.94$. The results are shown in Fig.~\ref{fig:panel_ic:q6_spch_b200}.

\bigskip
To conclude, the QML model can be trained with only a subset of Pauli measurement bases.

\section{Conclusion and outlook \label{sec:conclusion}}
To sum up, we have developed a method to use quantum machine learning to perform quantum state tomography. Our method only requires the classical measurements of the target state as the dataset, rather than the target state itself as the quantum data input. This makes our model amenable to NISQ devices and has enabled us to train and execute our model on quantum computers.
To illustrate the viability of our methodology, we use our model to reconstruct the GHZ state as well as the ground state of a spin chain Hamiltonian for the 3- and 6-qubit cases on ideal simulators. For the 6-qubit spin chain ground state, fidelity above $0.9$ can be obtained; for the other three cases, fidelity can reach $0.99$ most of the time. Additionally, quantum states produced by random circuits can be reconstructed, reaching $0.9$ fidelity typically and above 0.99 in some cases. Besides running on ideal simulators, we also verify the effectiveness of our model on actual quantum computers by training it on IBM Brisbane and IonQ Forte. The 3-qubit GHZ state can be reconstructed with fidelity above $0.9$ in both cases, demonstrating the possibility of implementation on NISQ devices. 
Additionally, we also attempt to reduce the required resource by training on incomplete measurement bases, which is an essential prerequisite for scaling to large systems. For this purpose, we perform QST with only a subset of Pauli bases on the GHZ state and spin chain ground state, using 15 out of 27 bases for the 3-qubit cases and 200 out of 729 bases for the 6-qubit cases, achieving fidelity over 0.9 for the 6-qubit spin chain state and over 0.98 for the other three cases.
In a nutshell, this study showcases the feasibility of using QML for QST with classical measurement data and lays the ground for future studies.

Several questions remain to be studied in the future.
First of all, as QML models may face a trainability issue known as the ``barren plateau'' problem when scaling up~\cite{thanasilp_subtleties_2023,rudolph_trainability_2023}, strategies to mitigate the barren plateau should be tested on tomography tasks of larger systems, including initialization schemes~\cite{grant_initialization_2019,kulshrestha_beinit_2022}, optimization strategies~\cite{ostaszewski_structure_2021,skolik_layerwise_2021,gharibyan_hierarchical_2023,nadori_line_2025}, choice of loss function~\cite{rudolph_trainability_2023,letcher_tight_2024}, etc.
Secondly, incorporation of additional \textit{a priori} information may further aid the state reconstruction. For example, future extensions could adopt symmetry-guided measurement schemes for physically constrained states (e.g., translationally invariant systems).
Furthermore, while this paper focuses on the pure state, it is worthwhile to generalize the approach here to mixed states.
Lastly, one insightful question to answer is what classes of quantum states can be practically reconstructed by such QML methods. This is relevant 
not only to QST itself but also to understanding the power of QML.

\begin{acknowledgments}
We gratefully acknowledge support from the University of Maryland’s National Quantum Laboratory (QLab), which provided funding and access to IonQ quantum hardware for this work. We also thank IBM Quantum for supporting this research through access to their quantum computing resources.

\end{acknowledgments}

\section*{Code availability}
The source code used in our study can be found in \url{https://github.com/yjh-bill/qml_for_qst}.

\appendix

\section{Optimization methods}
\label{sec:optimization}
\subsection{Introduction to optimization methods}

For the training of a variational quantum circuit, different optimization algorithms exist. 
One commonly-used algorithm in machine learning is the gradient descent method, where the gradient of the loss function $L$ with respect to the parameters $\vec{\theta}$, namely $\vec\nabla_{\vec \theta} L$, is calculated, and then the parameters are optimized along the direction of the gradient: $\vec \theta \leftarrow \vec \theta-\alpha \vec\nabla_{\vec \theta} L$. 
There are also improved methods based on gradient-descent, such as ADAM~\cite{kingma_adam_2017}, RMSProp~\cite{tieleman_lecture65_2012}, etc.
In classical machine learning, the calculation of the gradient is done by the backpropagation~\cite{rumelhart_learning_1986}, which makes use of the chain rule. For a variational quantum circuit, since the circuit is only measured at the end, such technique is generally not applicable. Alternative methods can be used to calculate the gradient. One such method is the finite difference method, where for each component $\theta_i$ of the parameter set $\vec{\theta}$, we have
$$
\frac{\partial L\left(\vec\theta\right)}{\partial \theta_i}\approx\frac{L\left(\vec\theta + \Delta\theta_i \hat{e}_i \right)-L\left(\vec\theta-\Delta\theta_i \hat{e}_i\right)}{2\Delta \theta_i},
$$
where $\hat{e}_i$ is the unit vector along the $i^\text{th}$ direction,
to approximate the gradient.
An alternative method is called parameter shift~\cite{mitarai_quantum_2018,schuld_evaluating_2019}, which can yield exact gradient for observables with certain forms instead of an approximate result. It is common to have observables with the form
$$
f\left(\vec\theta\right)=\bra{0}U^\dagger\left(\vec{\theta}\right) O U\left(\vec{\theta}\right)\ket{0},
$$
where $U\left(\vec{\theta}\right)=U_n(\theta_n) U_{n-1}(\theta_{n-1})...U_2 (\theta_2) U_1(\theta_1)$.  If each $U_i$ has the form 
$$U_i=\exp\left(-i \theta_i P_i/2\right),$$
where $P_i$ is a Pauli operator, then one has
$$
\frac{\partial f\left(\vec\theta\right)}{\partial \theta_i}=\frac{1}{2}\left(f\left(\vec\theta+\frac{\pi}{2}\hat{e}_i\right)-f\left(\vec\theta-\frac{\pi}{2}\hat{e}_i\right)\right).
$$
With these techniques, the gradient can be evaluated. However, if a model has $n$ parameters, then for each parameter $\theta_i$ ($i=1, 2, ..., n$), the quantum circuit needs to be evaluated twice. This means that $2n$ function calls need to be made for each iteration. Due to this overhead, the computation cost of gradient-based methods is often higher than some non-gradient based methods.

Among the non-gradient methods, one algorithm is the simultaneous perturbation stochastic approximation (SPSA) algorithm.  Rather than calculating the gradient of the loss function, SPSA chooses a random direction according to certain rules for each iteration, and the updating of parameters is based on the derivative of the loss function along this direction. 
More precisely, a random direction $\vec\Delta$ is chosen for each iteration. Then, a ``gradient'' is estimated according to the rule
$$
\vec{g}_k=\frac{L\left(\vec{\theta}+c_k \vec \Delta\right)-L\left(\vec{\theta}-c_k \vec \Delta\right)}{2 c_k}
\begin{pmatrix}\Delta_1^{-1} \\ \Delta_2^{-1} \\ \vdots \\ \Delta_n^{-1}\end{pmatrix},
$$
where $k$ is the number of iterations, $\Delta_i$ refers the $i^\text{th}$ element of $\vec{\Delta}$, and $c_k$ has the form $c_k=c/k^\gamma$. Then, the parameters $\vec\theta$ are updated according to 
$$
\vec{\theta}\leftarrow \vec{\theta}-a_k \vec g_k,
$$
where $a_k=a/(A+k)^\alpha$. Here, $c$, $\gamma$, $a$, $A$, and $\alpha$ are hyperparameters
\setcounter{footnote}{100}
\footnote{In the main text, for the 6-qubit case, the default hyperparameters are used. For the 3-qubit case, we have $c=0.1258$, $A=0.3186$, $a_1=0.4739$, $\alpha=0.6374$, and $\gamma=0.06059$.\label{footnote:hyperparameters}}.

Other gradient-free methods exist as well. In Python's Scipy library~\cite{2020SciPy-NMeth}, many standard optimization algorithms are provided, including the Powell algorithm, the COBYLA algorithm, etc.

\subsection{Comparison between optimization algorithms}

\newcommand{\scaleD}{0.9}
\begin{figure}[htbp]
    \begin{subfigure}[t]{\linewidth}
        \centering
        \includegraphics[width=\scaleD\linewidth]{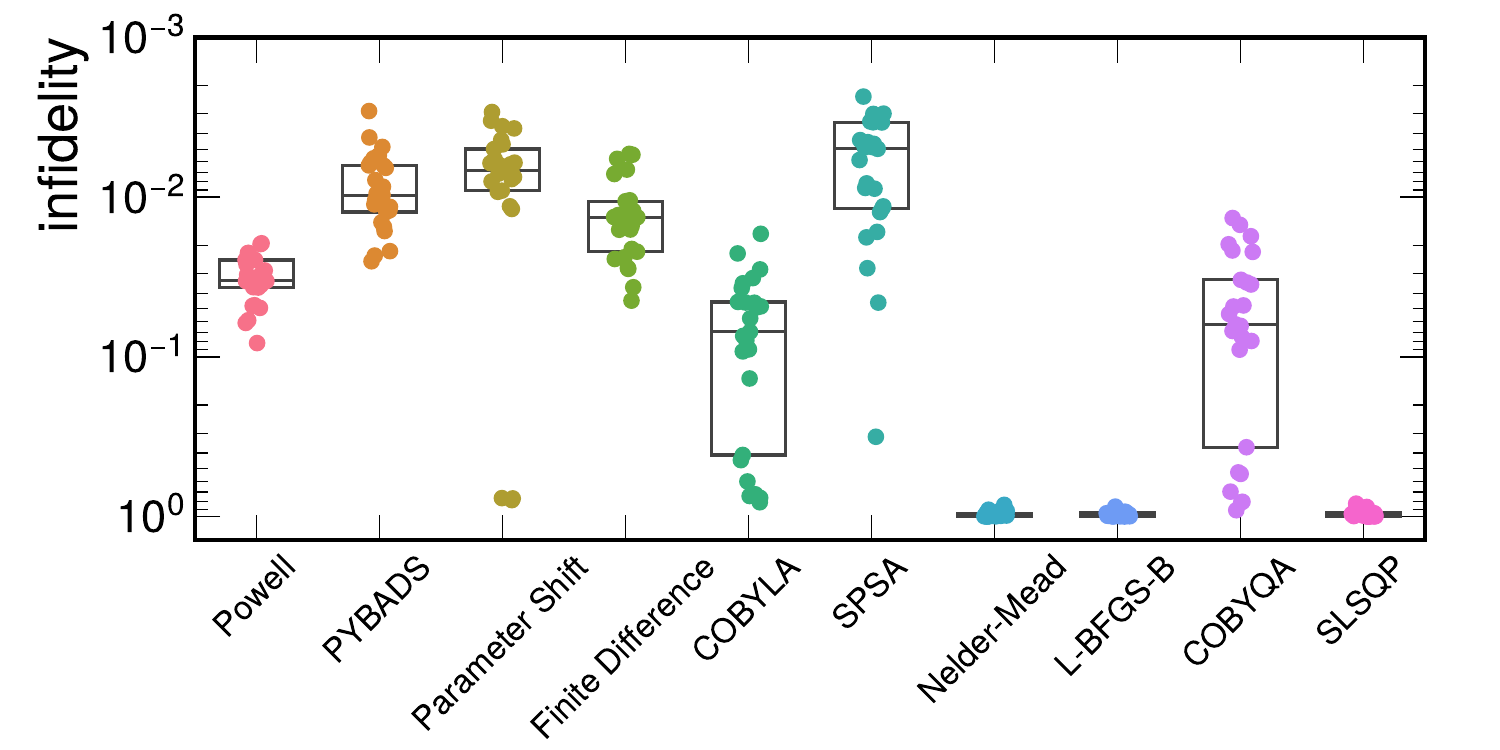}
        \caption{Infidelity}
        \label{fig:comparison_ghz_all:fidelity}
    \end{subfigure}

    \begin{subfigure}[t]{\linewidth}
        \centering
        \includegraphics[width=\scaleD\linewidth]{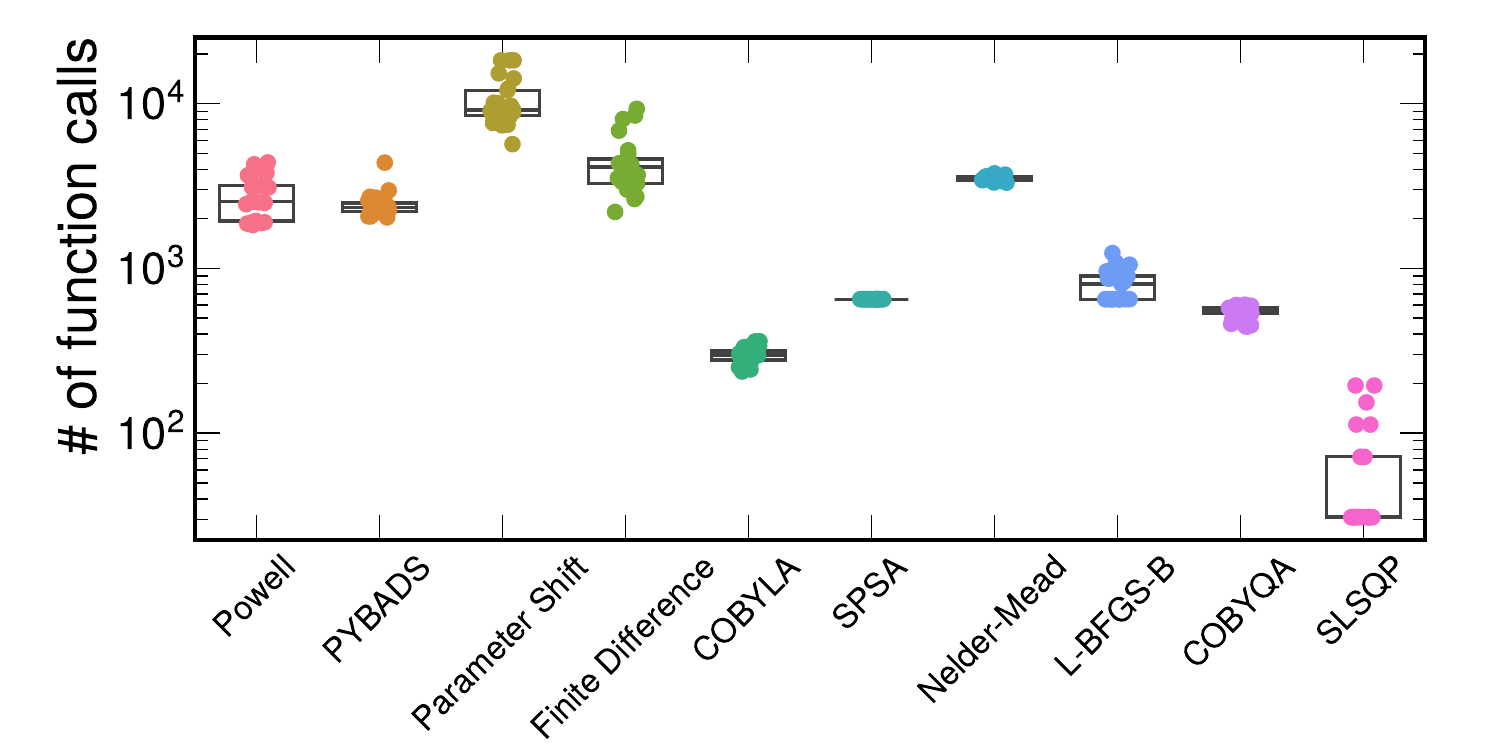}
        \caption{Computation cost}
        \label{fig:comparison_ghz_all:cost}
    \end{subfigure}
    
    \caption{Comparison across different optimization methods for the 3-qubit GHZ state. The obtained infidelities are shown in \ref{fig:comparison_ghz_all:fidelity}, while the numbers of function calls are shown in \ref{fig:comparison_ghz_all:cost}.
    }
    \label{fig:comparison_ghz_all}
\end{figure}

\begin{figure}[htbp]
    \begin{subfigure}[t]{\linewidth}
        \centering
        \includegraphics[width=\scaleD\linewidth]{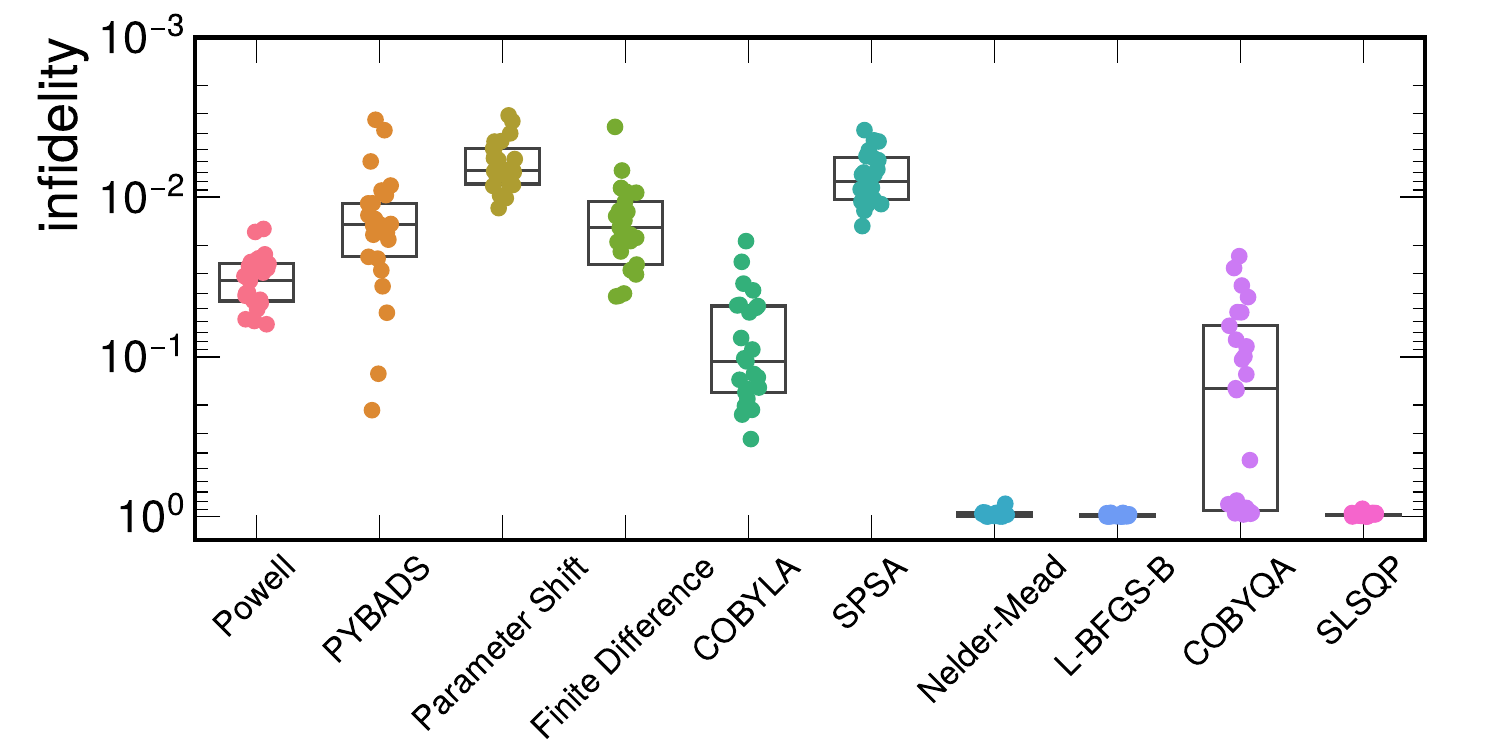}
        \caption{Infidelity}
        \label{fig:comparison_spin_chain_all:fidelity}
    \end{subfigure}

    \begin{subfigure}[t]{\linewidth}
        \centering
        \includegraphics[width=\scaleD\linewidth]{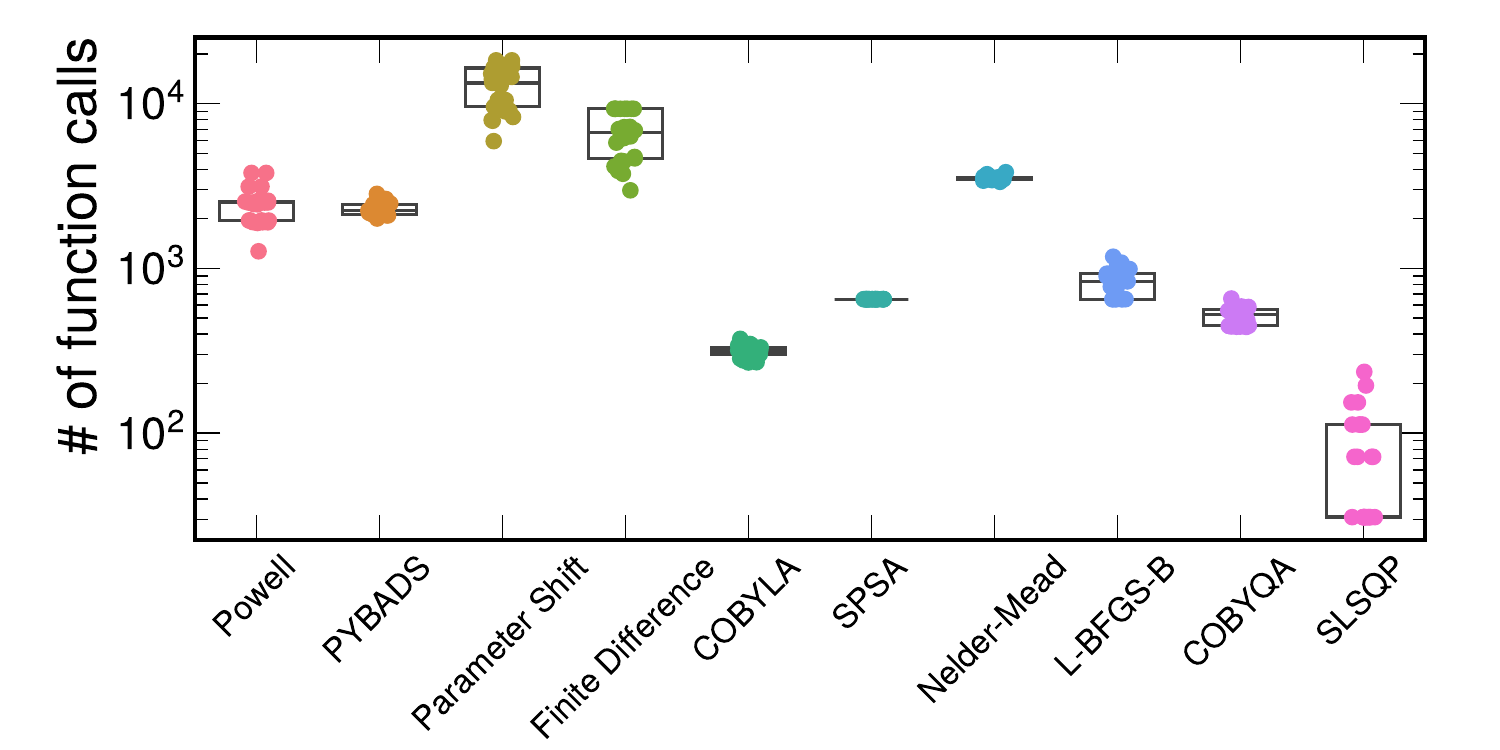}
        \caption{Computation cost}
        \label{fig:comparison_spin_chain_all:cost}
    \end{subfigure}

    \caption{Comparison across different optimization methods for the 3-qubit spin chain ground state. The obtained infidelities are shown in \ref{fig:comparison_spin_chain_all:fidelity}, while the numbers of function calls are shown in \ref{fig:comparison_spin_chain_all:cost}.
    }
    \label{fig:comparison_spin_chain_all}
\end{figure}

Here, we give a comparison between many different optimization methods. 
The tested methods include gradient-based methods, namely the finite difference method and the parameter shift method, as well as various gradient-free methods.
Here, the parameter shift method was implemented with the help of \verb|ParameterShiftSamplerGradient| in Qiskit~\cite{qiskit2024}, and the finite difference method was implemented manually~\footnote{The Adam optimizer~\cite{kingma_adam_2017} was used for both the parameter shift method and the finite difference method.}. For the SPSA algorithm, the implementation in Qiskit was used~\cite{qiskit2024}; the default hyperparameters were used here, unlike what was done in the main text~\cite{Note101}.
We used the `Pybads' python package for the Pybads optimizer~\cite{singh2024pybads,acerbi2017practical}.
For other optimizers, we used the implementation in \verb|scipy.optimize| \cite{2020SciPy-NMeth}.
The optimization algorithms were tested on the 3-qubit GHZ state and the 3-qubit spin chain ground state. 
The training settings were the same as in Section \ref{sec:results:ideal}.  The simulation was performed on the ideal simulator. 25 runs were performed for each optimization method. 
For the 3-qubit GHZ state, the results are shown in Fig.~\ref{fig:comparison_ghz_all}, with Fig.~\ref{fig:comparison_ghz_all:fidelity} showing the distribution of infidelity and ~\ref{fig:comparison_ghz_all:cost} showing the computation cost measured in terms of the number of function calls
\footnote{
Note, instead of the number of iterations during training, we compare the number of function calls instead. This is because different optimizers may require different numbers of function calls for each iteration. For example, in COBYLA, the function is called once for every iteration; in SPSA, the function is called twice for each iteration; in gradient-based methods, the function is called multiple times for each iteration. To take this into account, it makes more sense to use the number of function calls to measure the cost.
}.
The results for the 3-qubit spin chain ground state are similarly shown in Fig.~\ref{fig:comparison_spin_chain_all}.
Among the tested methods, Powell, Pybads, SPSA, COBYLA, and COBYQA perform well, with the typical fidelity reaching $0.9$, while the Nelder-Mead, L-BFGS-B, and SLSQP algorithms fail to train the model. 
Among the functioning ones, the COBYLA optimizer is the most efficient, requiring the fewest number of function calls to train, and the SPSA algorithm is the second most efficient in the test cases here. The Powell optimizer and Pybads optimizer take significantly more time to train. Gradient-based methods, namely the finite difference method and the parameter shift method, take the most time to train due to the cost of computing the gradient. In terms of fidelity, COBYLA is outperformed by other methods, while SPSA has one of the highest fidelity. Thus, based on the test results, SPSA is the top choice for our task.
Note, the performance and computation of these optimizers may vary under different hyperparameter settings. To search for the optimal setting for each optimizer is beyond the scope of this work. Instead, the comparison here is meant to give an intuitive picture of the pros and cons of various optimizers.

\section{Alternative loss function}
\label{sec:alt_loss}

\begin{figure}[hbt!]
    \centering
    \begin{subfigure}[t]{0.48\linewidth}
        \centering
        \includegraphics[width=\linewidth]{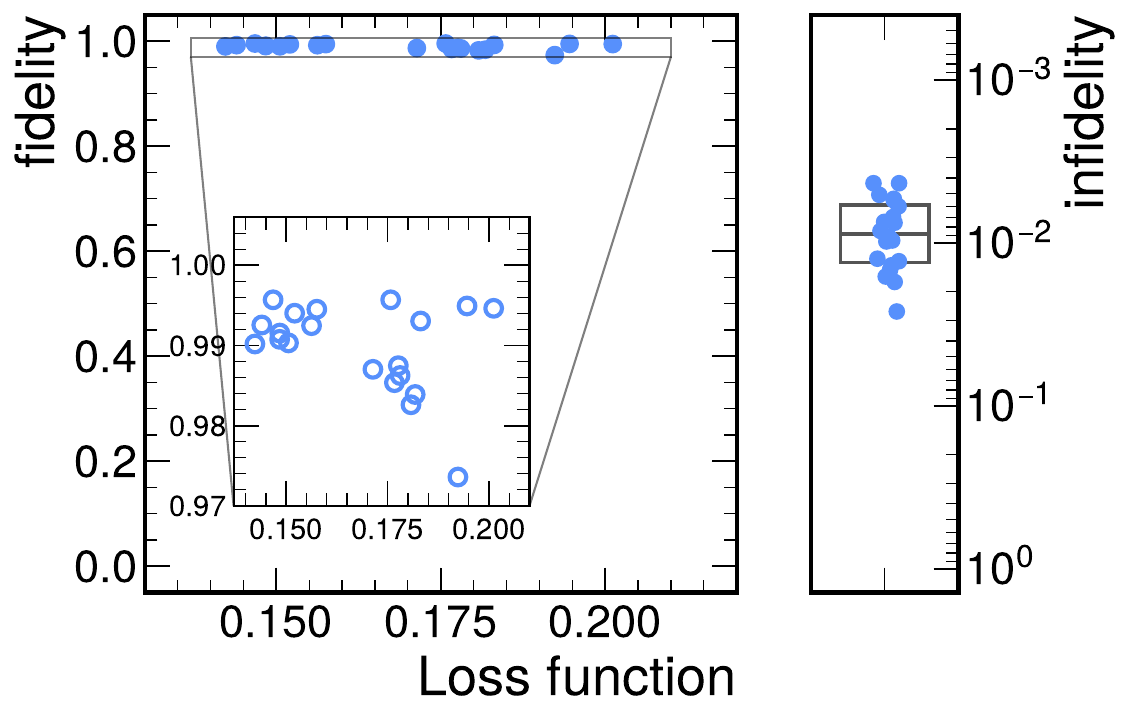}
        \caption{3-qubit GHZ state}
        \label{fig:panel_mmd:q3_ghz}
    \end{subfigure}
    \hspace{0.2em}
    \begin{subfigure}[t]{0.48\linewidth}
        \centering
        \captionsetup{justification=centering}
        \includegraphics[width=\linewidth]{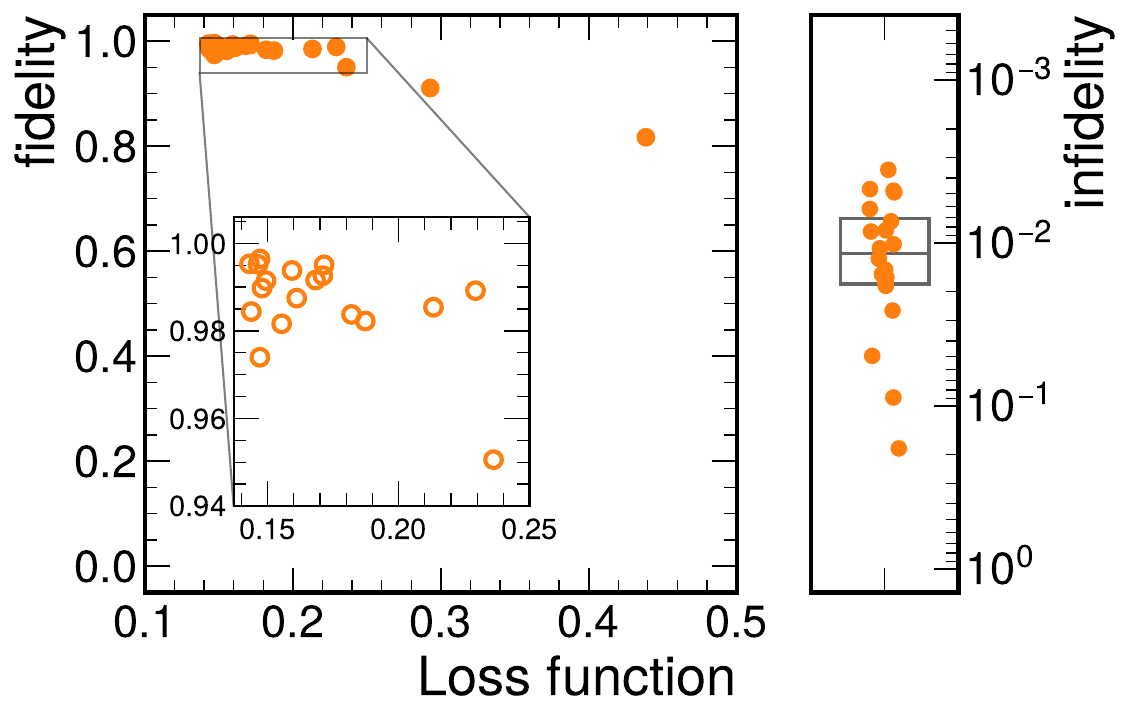}
        \caption{3-qubit spin chain ground state}
        \label{fig:panel_mmd:q3_spch}
    \end{subfigure}

    \begin{subfigure}[t]{0.48\linewidth}
        \centering
        \includegraphics[width=\linewidth]{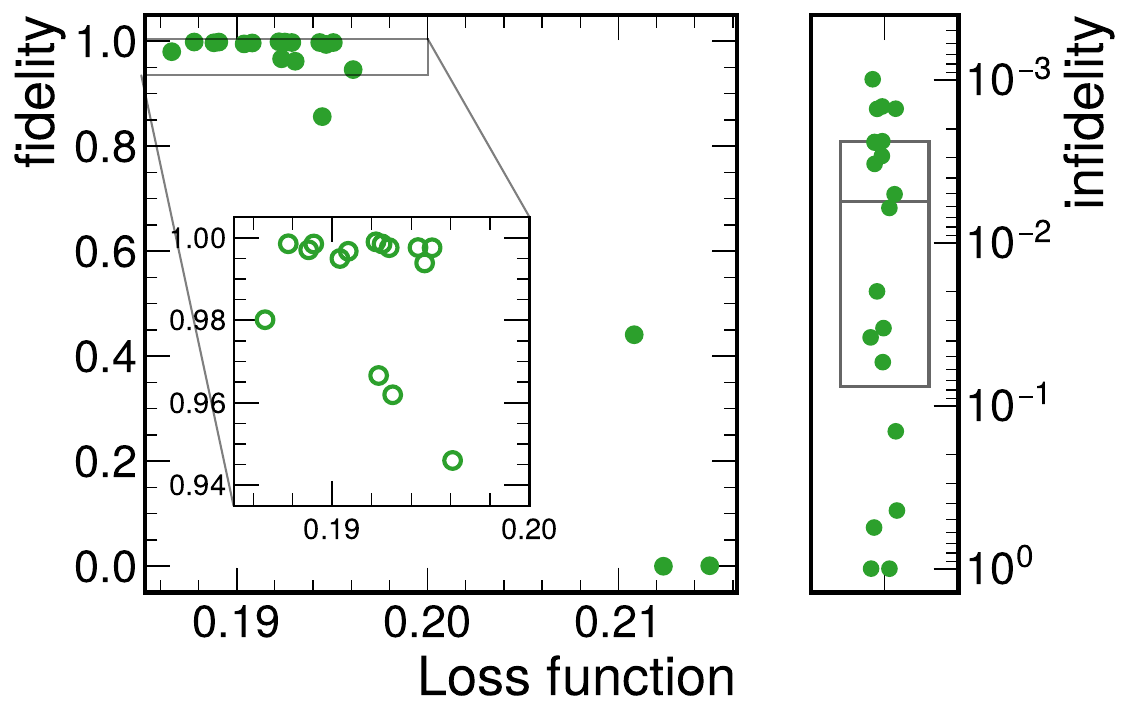}
        \caption{6-qubit GHZ state}
        \label{fig:panel_mmd:q6_ghz}
    \end{subfigure}
    \hspace{0.2em}
    \begin{subfigure}[t]{0.48\linewidth}
        \centering
        \captionsetup{justification=centering}
        \includegraphics[width=\linewidth]{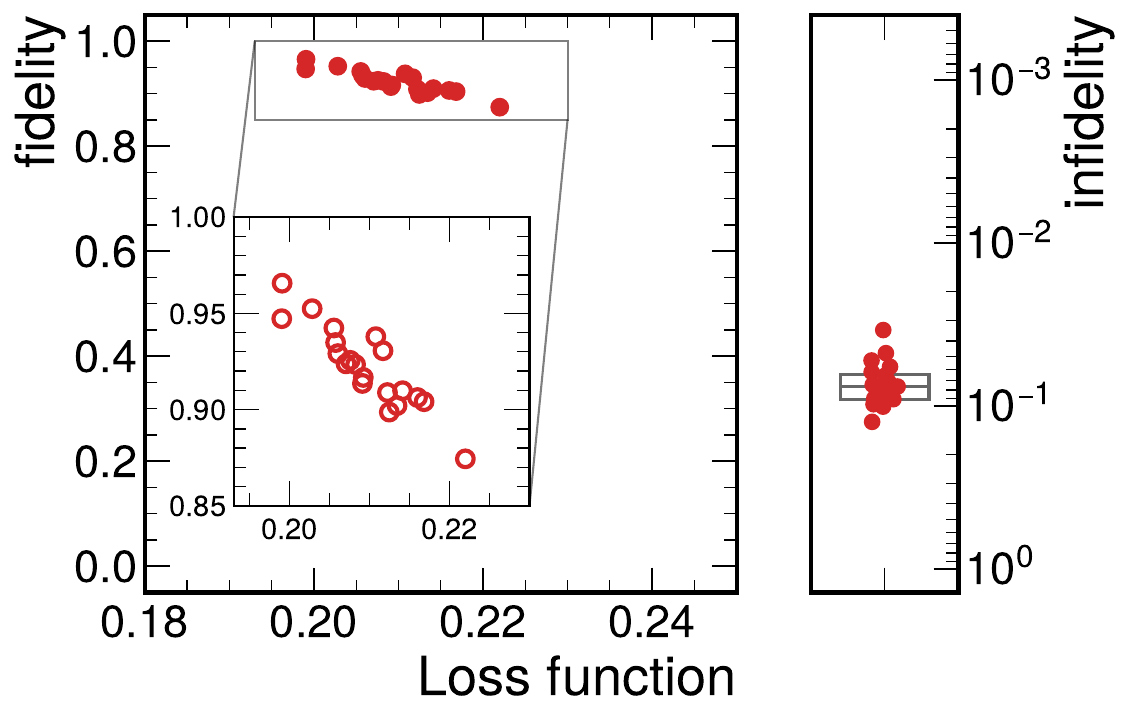}
        \caption{6-qubit spin chain ground state}
        \label{fig:panel_mmd:q6_spch}
    \end{subfigure}

    \caption{QST with MMD loss function. Other than the loss function, training settings are the same as Fig.~\ref{fig:panel}. In the plots shown here, the loss function is averaged across all bases and rescaled by a factor of 10.}
    \label{fig:panel_mmd}
\end{figure}

In the main text, we used the KL-divergence to measure the distance between any two distributions. However, alternative choices of loss functions exist. 
In particular, one possible choice is the maximum mean discrepancy (MMD) loss. Suppose one is given two probability distributions, $q_{(\theta)}$ and $p$, then the MMD between them is given by
\begin{align*}
L_{MMD}(\vec{\theta})=&\mathbb{E}_{\vec{x},\vec{y}\sim q_{\vec\theta}}\left[K\left(\vec x, \vec y\right)\right]
-2\mathbb{E}_{\vec{x}\sim q_{\vec\theta},y\sim p}\left[K\left(\vec x, \vec y\right)\right] \\
&+\mathbb{E}_{\vec{x},\vec{y}\sim p}\left[K\left(\vec x, \vec y\right)\right],
\end{align*}
where $K(\vec{x},\vec{y})$ is called the kernel function, and many function forms can be chosen as the kernel function. One commonly used kernel function is the Gaussian kernel:
\begin{equation}
K(\vec{x},\vec{y})=\exp\left(-\frac{\left\| \vec{x}-\vec{y}\right\|_2^2}{2\sigma}\right),
\label{eqn:mmd}
\end{equation}
where $\sigma$ is a hyperparameter called bandwidth. In Ref.~\cite{rudolph_trainability_2023}, it is conjectured that the MMD loss may alleviate the barren plateau problem under certain conditions. For this reason, we tested the performance of the MMD loss function on the GHZ state and the spin chain ground state. Here, we present the results with parameter $\sigma$ in Equation~\ref{eqn:mmd} fixed to $0.1$. The results are shown in Fig.~\ref{fig:panel_mmd}.  For the 3-qubit GHZ state, fidelity is above $0.98$ in most trials, with the median being $0.991$ in 20 trials. Similar performance is achieved for the 3-qubit spin chain ground state. Except for a few outliers, the fidelity is also typically above 0.98, and the median is $0.988$. 
For the 6-qubit GHZ, there exist a few outliers with low fidelity. However, excluding them, the other trials have fidelity above 0.9 and many achieve fidelity above 0.99. The median fidelity is $0.994$.
As for the 6-qubit spin chain ground state, fidelity near or above 0.9 can be achieved, with the median fidelity being $0.92$.
The results here demonstrate that the MMD loss can act as the loss function for quantum state tomography. 
Nonetheless, for these test cases, the performance of MMD does not exceed that of the KL-divergence.
Whether the MMD loss will exhibit an advantage on larger systems remains to be studied in the future.

\section{Reconstruction of random circuits \label{sec:random_circuits}}
Here, we show the random circuits that we use as target states in Section~\ref{sec:results:ideal}. See Fig.~\ref{fig:random_q3_circuits} and Fig.~\ref{fig:random_q6_circuits}.

\newcommand{\scaleB}{1.0}
\newcommand{\scaleBa}{0.03}
\newcommand{\scaleBb}{0.9}

\begin{figure}[htb!]
    \centering
    \begin{subfigure}[t]{\scaleB\linewidth}
        \begin{minipage}[c]{\scaleBa\linewidth}
            \caption{}
        \end{minipage}
        \begin{minipage}[c]{\scaleBb\linewidth}
            \includegraphics[width=\linewidth]{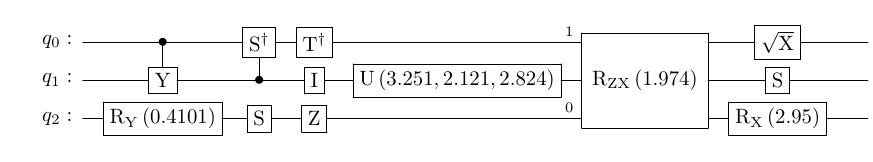}
        \end{minipage}
    \end{subfigure}

    \begin{subfigure}[t]{\scaleB\linewidth}
        \begin{minipage}[c]{\scaleBa\linewidth}
            \caption{}
        \end{minipage}
        \begin{minipage}[c]{\scaleBb\linewidth}
            \includegraphics[width=\linewidth]{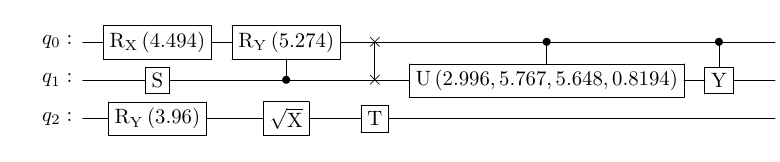}
        \end{minipage}
    \end{subfigure}

    \begin{subfigure}[t]{\scaleB\linewidth}
        \begin{minipage}[c]{\scaleBa\linewidth}
            \caption{}
        \end{minipage}
        \begin{minipage}[c]{\scaleBb\linewidth}
            \includegraphics[width=\linewidth]{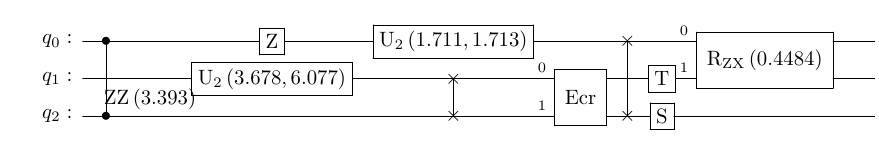}
        \end{minipage}
    \end{subfigure}

    \begin{subfigure}[t]{\scaleB\linewidth}
        \begin{minipage}[c]{\scaleBa\linewidth}
            \caption{}
        \end{minipage}
        \begin{minipage}[c]{\scaleBb\linewidth}
            \includegraphics[width=\linewidth]{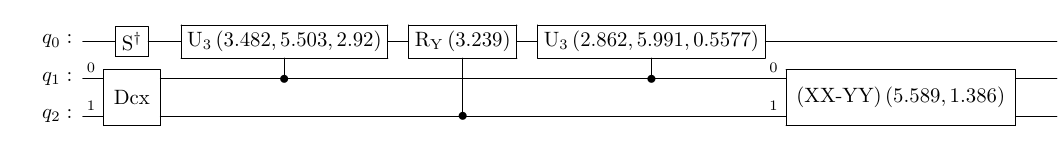}
        \end{minipage}
    \end{subfigure}

    \begin{subfigure}[t]{\scaleB\linewidth}
        \begin{minipage}[c]{\scaleBa\linewidth}
            \caption{}
        \end{minipage}
        \begin{minipage}[c]{\scaleBb\linewidth}
            \includegraphics[width=\linewidth]{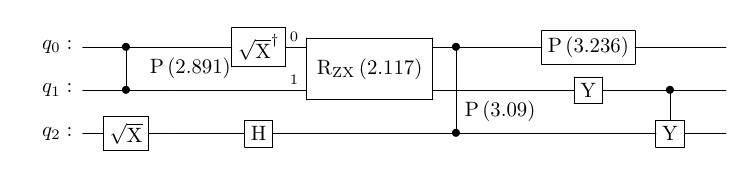}
        \end{minipage}
    \end{subfigure}
    
    \caption{Random 3-qubit circuits}
    \label{fig:random_q3_circuits}
\end{figure}

\newcommand{\scaleC}{1.0}
\newcommand{\scaleCa}{0.03}
\newcommand{\scaleCb}{0.9}

\begin{figure}[htb!]
    \centering
    \begin{subfigure}[t]{\scaleC\linewidth}
        \begin{minipage}[c]{\scaleCa\linewidth}
            \caption{}
        \end{minipage}
        \begin{minipage}[c]{\scaleCb\linewidth}
            \includegraphics[width=\linewidth]{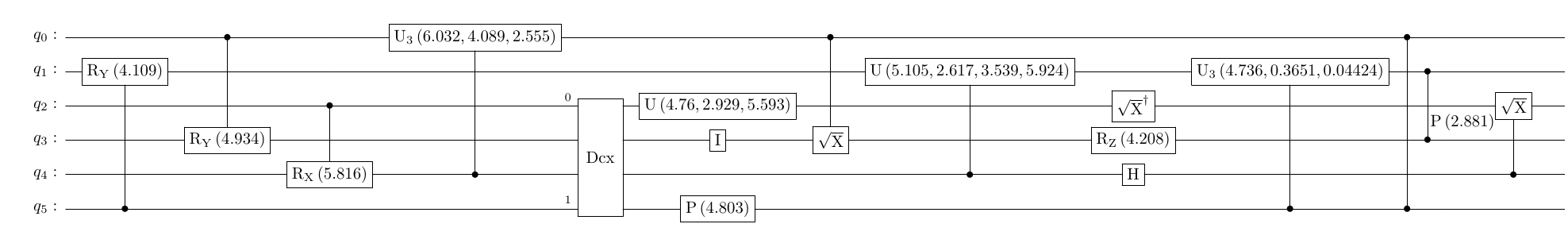}
        \end{minipage}
    \end{subfigure}

    \begin{subfigure}[t]{\scaleC\linewidth}
        \begin{minipage}[c]{\scaleCa\linewidth}
            \caption{}
        \end{minipage}
        \begin{minipage}[c]{\scaleCb\linewidth}
            \includegraphics[width=\linewidth]{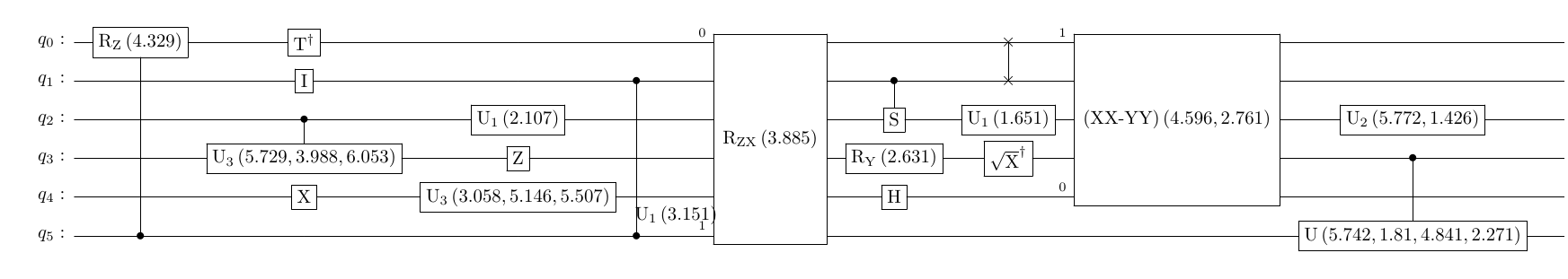}
        \end{minipage}
    \end{subfigure}

    \begin{subfigure}[t]{\scaleC\linewidth}
        \begin{minipage}[c]{\scaleCa\linewidth}
            \caption{}
        \end{minipage}
        \begin{minipage}[c]{\scaleCb\linewidth}
            \includegraphics[width=\linewidth]{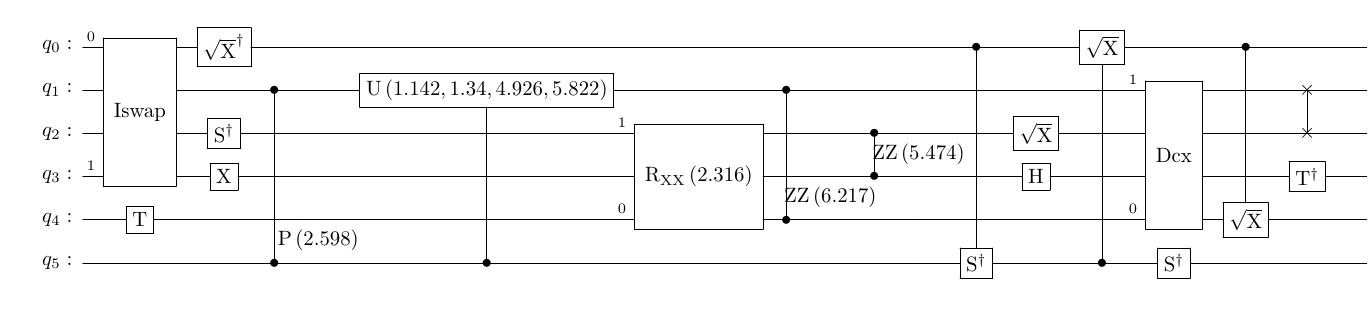}
        \end{minipage}
    \end{subfigure}

    \begin{subfigure}[t]{\scaleC\linewidth}
        \begin{minipage}[c]{\scaleCa\linewidth}
            \caption{}
        \end{minipage}
        \begin{minipage}[c]{\scaleCb\linewidth}
            \includegraphics[width=\linewidth]{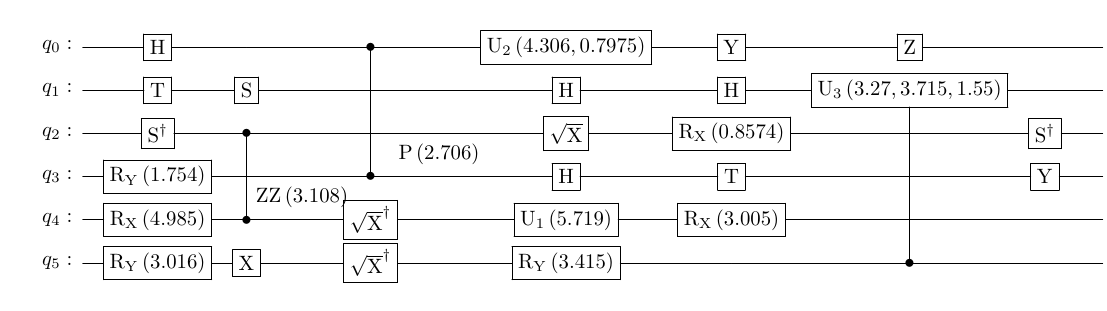}
        \end{minipage}
    \end{subfigure}

    \begin{subfigure}[t]{\scaleC\linewidth}
        \begin{minipage}[c]{\scaleCa\linewidth}
            \caption{}
        \end{minipage}
        \begin{minipage}[c]{\scaleCb\linewidth}
            \includegraphics[width=\linewidth]{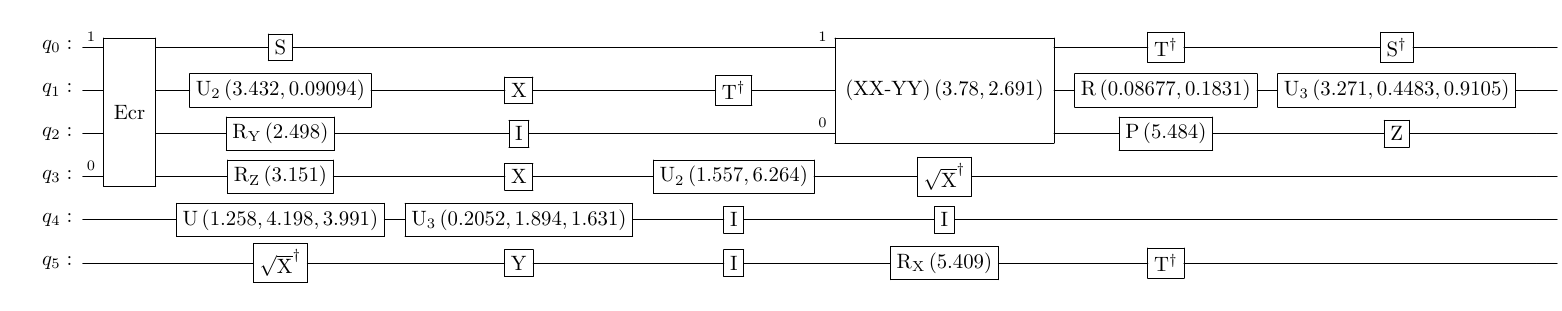}
        \end{minipage}
    \end{subfigure}
    
    \caption{Random 6-qubit circuits}
    \label{fig:random_q6_circuits}
\end{figure}

\FloatBarrier
\bibliography{refs.bib}

\end{document}